\documentclass{aa}

\usepackage{lipsum}
\usepackage{graphicx}
\usepackage{txfonts}
\usepackage{xcolor}
\usepackage{orcidlink}
\usepackage{microtype}
\usepackage{xspace}

\usepackage{siunitx}
\DeclareSIUnit\parsec{pc}
\DeclareSIUnit\solarmass{\textit{M}\textsubscript{\(\odot\)}}
\DeclareSIUnit\gauss{G}
\DeclareSIUnit\erg{erg}
\DeclareSIUnit\year{yr}
\DeclareSIUnit\deg{deg}
\DeclareSIUnit\arcmin{arcmin}
\DeclareSIUnit\arcsec{arcsec}
\DeclareSIUnit\counts{cts}
\DeclareSIUnit\dex{dex}
\DeclareSIUnit\jansky{Jy}
\DeclareSIUnit\beam{beam}
\DeclareSIUnit\steradian{sr}
\DeclareSIUnit\hertz{Hz}

\usepackage{ctable}
\usepackage{hyperref}
\hypersetup{    breaklinks=true,
                colorlinks=true,
                citecolor=blue,
                linkcolor=blue,
                urlcolor=blue}

\newcommand{\hi}{H\,{\sc i}\xspace}
\newcommand{\mathhi}{\textsc{H\,i}}
\newcommand{\hicap}{{{\rm H}\,{\tiny\rm I}}}

\newcommand{\bbarolo}{\textsuperscript{3D}BAROLO}

\begin{document}

    \title{H\,{\Large I} rotation curves and mass-size relation for 568 galaxies from the Apertif Wide-area Extragalactic Survey}

    \titlerunning{H\,{\fontsize{8}{0}\selectfont{}I} rotation curves and mass-size relation for 568 galaxies from AWES}

    \author{
            Konstantin~Haubner\inst{\ref{inst1},\ref{inst2}}\textsuperscript{,}\thanks{Corresponding author: \texttt{konstantin.haubner@inaf.it}}\orcidlink{0009-0007-7808-4653}
        \and
            Federico~Lelli\inst{\ref{inst1}}\orcidlink{0000-0002-9024-9883}
        \and
            Enrico Di~Teodoro\inst{\ref{inst2},\ref{inst1}}\orcidlink{0000-0003-4019-0673}          
        \and
            Kelley M. Hess\inst{\ref{inst3},\ref{inst4}, \ref{inst5}}\orcidlink{0000-0001-9662-9089}
        \and                   
            Francis~Duey\inst{\ref{inst6}}\orcidlink{0009-0003-1662-5179}
        \and
            Stacy~McGaugh\inst{\ref{inst6}}\orcidlink{0000-0002-9762-0980}
        \and
            James~Schombert\inst{\ref{inst7}}\orcidlink{0000-0003-2022-1911}
        \and
            Elizabeth A. K. Adams\inst{\ref{inst4},\ref{inst8},\ref{inst9}}\orcidlink{0000-0002-9798-5111}
        \and
            W.J.G. de Blok\inst{\ref{inst4}, \ref{inst10}, \ref{inst9}}\orcidlink{0000-0001-8957-4518}
        \and
            Danielle M. Lucero\inst{\ref{inst11}}\orcidlink{0000-0002-9288-9331}
        \and
            Helga Dénes\inst{\ref{inst12}}\orcidlink{0000-0002-9214-8613}
        \and
            Julián Garrido\inst{\ref{inst5}}\orcidlink{0000-0002-6696-4772}
        \and
            Joeri van Leeuwen\inst{\ref{inst4}}\orcidlink{0000-0001-8503-6958}
        \and
            Manuel Parra-Royón\inst{\ref{inst5}}\orcidlink{0000-0002-6275-8242}
        \and
            Anastasia A. Ponomareva\inst{\ref{inst13}}\orcidlink{0000-0003-4100-0173}
        \and
            Susana Sánchez Exposito\inst{\ref{inst5}}\orcidlink{0000-0002-7510-7633}
}

    \institute{
            INAF -- Arcetri Astrophysical Observatory, Largo Enrico Fermi 5, 50125 Firenze, Italy\label{inst1}
        \and
            Dipartimento di Fisica e Astronomia, Università degli Studi di Firenze, via G. Sansone 1, 50019 Sesto Fiorentino, Firenze, Italy\label{inst2}
        \and
            Department of Space, Earth and Environment, Chalmers University of Technology, Onsala Space Observatory, 43992 Onsala, Sweden\label{inst3}
        \and
            Netherlands Institute for Radio Astronomy (ASTRON), Oude Hoogeveensedijk 4, 7991 PD Dwingeloo, The Netherlands\label{inst4}
        \and
            Instituto de Astrofísica de Andalucía (CSIC), Glorieta de la Astronomía s/n, 18008 Granada, Spain\label{inst5}
        \and
            Department of Astronomy, Case Western Reserve University, Cleveland, OH 44106, USA\label{inst6}
        \and
            Department of Physics, University of Oregon, Eugene, OR 97403, USA\label{inst7}
        \and
            SKA Observatory, SKA-Mid Science Operations Centre, 2 Fir Street, Observatory Cape Town, 7925 South Africa\label{inst8}
        \and
            Kapteyn Astronomical Institute, University of Groningen, PO Box 800, 9700 AV Groningen, The Netherlands\label{inst9}
        \and
            Department of Astronomy, University of Cape Town, Private Bag X3, Rondebosch 7701, South Africa\label{inst10}
        \and
            Department of Physics, Virginia Tech, 850 West Campus Drive, Blacksburg, VA 24061, USA\label{inst11}
        \and
            College of Sciences and Engineering, Universidad San Francisco de Quito, Quito, 170901, Ecuador\label{inst12}
        \and
            Centre for Astrophysics Research, School of Physics, Astronomy and Mathematics, University of Hertfordshire, College Lane, Hatfield, AL10 9AB, UK\label{inst13}
}

    \date{Received June 5, 2026; accepted August 20, 2026}

    \abstract{
        The extended \hi rotation curves of galaxies are one of the most powerful tools for studying galaxy evolution and for testing different dark matter and modified gravity models. However, their sample sizes have historically been limited to a few hundred galaxies, mostly due to the survey speed of radio interferometers. Here, we present a first data release of \hi kinematic models from the Apertif Wide-area Extragalactic Survey (AWES). Starting from 1924 galaxies detected in \hi, we constructed 568 reliable kinematic models, which provide rotation curves, velocity dispersion profiles, surface density profiles, and \hi disk geometries. This dataset constitutes the largest homogeneous \hi rotation curve sample to date. The \hi disks of the modeled galaxies span three orders of magnitude in \hi mass, have outer rotation velocities between ${\sim}20$ and ${\sim}\SI{300}{\kilo\meter\per\second}$, and extend from 2 to more than \SI{60}{\kilo\parsec}. The number of data points of the rotation curves ranges from 3 to 23, with a peak value at 4 and a broad tail to larger values. The rotation curves display a variety of inner shapes, but in general, they tend to become flat at large radii. Based on the \hi surface density profiles derived during the kinematic modeling, we studied the \hi mass-size relation with a large homogeneous sample of 213 galaxies. The relation is very tight, with an observed scatter along the mass direction of ${\sim}\SI{0.101}{\dex}$ and an inferred intrinsic scatter along the same direction of only \SI{0.067\pm 0.007}{\dex}. The best fitting slope is $1.96\pm0.03$ when fitting for mass as a function of size, indicating that \hi disks have a constant average surface density.
    }

    \keywords{
        astronomical data bases: miscellaneous --
        galaxies: kinematics and dynamics -- galaxies: ISM --
        galaxies: structure --
        cosmology: dark matter --
        radio lines: galaxies                   }

    \maketitle

\nolinenumbers

\section{Introduction}

Observations of the \SI{21}{\centi\meter} hyperfine structure line of neutral atomic hydrogen (\hi) are a crucial tool for studying many different aspects of the formation and evolution of galaxies. This includes the processes of gas accretion and star formation that regulate their overall growth across cosmic times, their interactions with other galaxies and the surrounding environment, and their internal dynamics. In particular, \hi typically forms a regularly rotating disk that allows tracing the kinematics of a galaxy well beyond its stellar component \citep[e.g.,][]{1978PhDT.......195Bosma, 1987PhDT.......199Begeman}. Therefore, it can be used to measure gravitational potentials and angular momentum out to large radii, where the discrepancy between dynamical and luminous masses becomes significant.

The information provided by \hi data allows us to study a range of  different scaling relations, which provide insights into the physics governing the dynamics and the evolution of galaxies. These include the baryonic Tully-Fisher relation \citep[BTFR;][]{2000ApJ...533L..99McGaugh, 2001ApJ...563..694Verheijen, 2016ApJ...816L..14Lelli, 2018MNRAS.474.4366P} and the radial acceleration relation \citep[RAR;][]{2016PhRvL.117t1101McGaugh, 2017ApJ...836..152Lelli, 2025MNRAS.541.2366Varasteanu}, which point to a significant coupling between the baryonic masses of galaxies and their dynamics. With respect to the structure of \hi disks, the key correlation is the \hi mass-size relation, which is a tight correlation between the \hi diameters, $D_\mathhi$, and the total \hi masses, $M_\mathhi$, of galaxies, taking the form of a power law \citep[e.g.,][]{1997A&A...324..877Broeils, 2002A&A...390..829Swaters, 2016MNRAS.460.2143Wang, 2022MNRAS.512.2697Rajohnson, 2024ApJ...976..159Deg}. Due to the sensitivity limits of historical radio interferometers, $D_\mathhi$ is typically defined as the diameter at an azimuthally averaged surface density of \SI{1}{\solarmass\per\square\parsec}. The small scatter of the \hi mass-size relation, along with its slope of almost exactly $\alpha = 2$ (when writing $M_\mathhi\propto {D_\mathhi}^\alpha$), indicates that galactic \hi disks have an almost constant average surface density. Recently, \citet{2025ApJ...980...25Wang} used very deep \hi data to demonstrate that the relation holds with an identical slope even when the diameter is defined at \SI{0.01}{\solarmass\per\square\parsec}. They interpret this as evidence for self-similar \hi surface density profiles in the outer regions of galaxies \citep[for a theoretical explanation see also][]{2019MNRAS.490...96Stevens}.

The tightness of the \hi mass-size relation makes it a powerful diagnostic for galaxy evolution models \citep[e.g., to identify feedback prescriptions that are too strong;][]{2016MNRAS.456.1115Bahe}, and for investigating the properties of \hi selected galaxy samples \citep[e.g.,][]{2016MNRAS.463.4052Ponomareva}. However, there is no unanimous agreement on the exact slope as some studies find shallower values \citep[$1.8-1.9$, e.g.,][]{2001A&A...370..765Verheijen, 2005A&A...442..137Noordermeer, 2016AJ....152..157Lelli, 2024ApJ...976..159Deg}, suggesting that the surface density of \hi disks might vary with mass. In addition, the intrinsic scatter of the relation has, to our knowledge, only been measured once \citep{2022MNRAS.512.2697Rajohnson}. An accurate measurement of the intrinsic scatter is crucial for interpreting the relation, as it reflects the intrinsic variation of \hi surface density profiles at fixed \hi mass.

The study of the kinematics and distribution of \hi, including the \hi mass-size relation, requires spatially resolved observations with radio interferometers. Due to relatively small fields of view (FoVs) and low array filling factors, most interferometric \hi surveys have been characterized by a low survey speed and most resolved \hi samples have only reached sizes of a few dozen to a few hundred galaxies. As a result, the statistical power of \hi dynamical and morphological studies has been limited. For example, this can be an issue when trying to measure the dependence of galaxy kinematics on the surrounding environment \citep{2020ApJ...904...51Chae, 2023MNRAS.519..318Kim} or the resolved galaxy velocity function \citep[see, e.g.,][]{2015MNRAS.454.1798Klypin}. These shortcomings can be addressed with better statistics, which could also improve measurements of the intrinsic scatter of scaling relations such as the \hi mass-size relation, BTFR, and RAR. The intrinsic scatter measured for the latter two relations is $\lesssim \SI{0.13}{\dex}$ and is currently consistent with zero \citep{2016ApJ...816L..14Lelli, 2017ApJ...836..152Lelli, 2018MNRAS.474.4366P}.

Luckily, the situation is improving due to pathfinders and precursors to the SKA Project. Two prime examples are the South African MeerKAT telescope and the Australian SKA Pathfinder (ASKAP). MeerKAT has a larger FoV and higher sensitivity than previous interferometers and allows for rapid \hi surveys of the sky \citep[e.g.,][]{2021A&A...646A..35Maddox, 2023A&A...673A.146Serra}, while ASKAP makes use of phased array feeds (PAFs). These allow for a greatly increased instantaneous FoV thanks to the simultaneous formation of 36 compound beams. ASKAP is currently conducting the Widefield ASKAP L-band Legacy All-sky Blind Survey \citep[WALLABY;][]{2020Ap&SS.365..118Koribalski}, which is planned to observe half of the southern sky in \hi. The pilot data releases~1 and 2 of WALLABY have been accompanied by the publication of more than 200 kinematic models \citep{2022PASA...39...59Deg, 2024PASA...41...88Murugeshan}, with an expected final sample size of thousands of kinematic models. Within the next decade, these numbers are expected to be surpassed by another order of magnitude by the SKA Project and the Deep Synoptic Array \citep[DSA;][]{2019BAAS...51g.255Hallinan}.

In this paper, we focus on \hi data taken with the SKA pathfinder Aperture Tile in Focus \citep[Apertif;][]{2019NatAs...3..188Adams, 2022A&A...658A.146VanCappellen}. Similarly to ASKAP, Apertif is a PAF system on 12 of the 14 Westerbork Synthesis Radio Telescope (WSRT) dishes. The simultaneous formation of 40 compound beams increases the array's instantaneous FoV from ${\sim}\SI{0.2}{\square\deg}$ to ${\sim}\SI{8}{\square\deg}$. The resulting upgrade in survey speed enabled two imaging surveys \citep{2022A&A...667A..38Adams}: the Apertif Wide-area Extragalactic Survey (AWES) and the Apertif Medium-deep Survey (AMDS). These were conducted between July 1, 2019 and February 28, 2022, with each individual \SI{11.5}{\hour} AWES and AMDS observation taking spectral line, continuum, and polarization data. AWES covers an area of ${\sim}\SI{2200}{\square\deg}$ with typically one and in some cases two observation per field. AMDS covers a smaller area of ${\sim}\SI{130}{\square\deg}$ with typically ten observations per field (and slightly less in some cases). The vast amount of data from AWES and AMDS allows for statistically powerful studies of the resolved \hi properties of galaxies and their resolved kinematics and dynamics. As a first step in this direction, we present kinematic models, including \hi rotation curves, that we derived for 568 galaxies observed with AWES.

Section~\ref{sec_data} of the paper describes the AWES data that we used, some preliminary sample cuts, the assignment of distances, and the calculation of \hi masses and diameters. Furthermore, it gives an overview of some basic properties of our galaxy sample. We present the kinematic modeling pipeline in Sect.~\ref{sec_kin}, including the treatment of edge-on galaxies, corrections for pressure support, and the estimation of uncertainties. The success rate of our pipeline in producing reliable kinematic models is discussed in Sect.~\ref{sec_results}, where we also present the resulting kinematic models and investigate the AWES \hi mass-size relation. We summarize our results in Sect.~\ref{sec_summary}.

\section{Data}  \label{sec_data}

\subsection{Apertif Wide-area Extragalactic Survey}

The spectral range of the AWES data ($1292.5 - \SI{1429.3}{\mega\hertz}$) was divided into four subsets (Cube0 to Cube3) during data processing \citep{2022A&C....3800514Adebahr, 2022A&A...667A..38Adams}. Our kinematic models are based on Cube2 with a topocentric spectral range from $1373.8$ to \SI{1418.4}{\mega\hertz}, corresponding to optical velocities from $424$ to \SI{10170}{\kilo\meter\per\second}. The channel width is \SI{36.6}{\kilo\hertz}, corresponding to \SI{7.7}{\kilo\meter\per\second}. Cube2 data were the initial target for source finding with the Source Finding Application~2 \citep[SOFIA~2;][]{2015MNRAS.448.1922Serra, 2021MNRAS.506.3962Westmeier}. An overview of the source finding is given in Sect.~2.3 of \citet{2025A&A...703A.295Siljeg}, while the detailed procedure will be described in a forthcoming publication (Hess et al. in prep.). At the time of conducting this study, source finding had been completed for an area that covers ${\sim}42\%$ of the Apertif footprint. In total, this provided us with 5954 data cubelets, which were cut out by SOFIA around the \hi detections.

The noise level around the galaxies in the SOFIA~2 cubes is $1.2_{-0.2}^{+0.6}\,\SI{}{\milli\jansky\per\beam}$ (median with 16th and 84th percentile), corresponding to a $3\sigma$ \hi column density sensitivity of \SI{1.7e20}{\per\square\centi\meter} over three channels (\SI{23.1}{\kilo\meter\per\second}) and for a typical angular resolution of $15''\times 24''$. In general, the spatial resolution is $15''\times15''/\sin{\delta}$, where $\delta$ is the declination of the pointing. However, for some observations, the restoring beams can vary significantly between different compound beams across the FoV (e.g., because of differences in baseline flagging). The AWES cubelets were produced after mosaicking the 40 compound beams in an Apertif field to achieve full survey depth before source finding. Consequently, the effective restoring beam is a combination of the contributions of the different compound beams. We calculate the effective restoring beam for each detection as a Gaussian whose parameters are the weighted averages of the parameters of the individual restoring beams that contributed to the cleaning in the final mosaic. The individual restoring beam weights are determined by the primary beam correction for each compound beam \citep{2022A&A...667A..40Denes} at the location of the source (Hess et al. in prep.). Furthermore, the AWES cubelets of well-resolved galaxies are characterized by an oversubtracted continuum emission (described in detail in Hess et al. in prep.). We corrected the affected cubelets as described in Appendix~\ref{app_baseline}.

\subsection{Mask creation and crossmatching}   \label{sec_prep}

We used the SOFIA data cubelets to create moment maps and masks, which \bbarolo\ needs to prevent low signal-to-noise (S/N) voxels from biasing the kinematic modeling results. Six of the SOFIA data cubelets contained bad channels, which we flagged. These were probably the result of imperfect continuum subtraction or radio frequency interference \citep{2022A&A...667A..38Adams}. We smoothed all cubelets to a minimum spatial resolution of $30''\times 30''$ with the \bbarolo\ task \texttt{SMOOTH}. We left the beam major axis unchanged if it was already larger than \SI{30}{\arcsec}. The smoothing is a necessary step because initial tests demonstrated that the S/N of most full-resolution AWES cubelets is too low for reliable kinematic modeling. This step improves the column density sensitivity to \SI{7.8e19}{\per\square\centi\meter} compared to the value for the full-resolution cubelets (\SI{1.7e20}{\per\square\centi\meter}).

To derive masks for the kinematic modeling, we used the \bbarolo\ task \texttt{SEARCH} with the \texttt{ITERNOISE} option to allow for an iterative estimation of the cubelet noise\footnote{We ran the code in single-thread mode, as the multi-thread mode can introduce some randomness and therefore decrease the reproducibility of the results.}. The \texttt{SEARCH} task produces a mask by first selecting all voxels above a primary S/N threshold and then extending the mask within each channel down to a secondary S/N threshold. The individual channel-based masks are merged if they are not separated by more than a chosen number of channels. After testing different threshold values, we chose a primary S/N threshold of 3.5, a secondary growth threshold of 2.5, and a velocity separation of two channels for merging the channel-based detections. Apart from producing the masks necessary for the kinematic modeling, this step excluded 3203 SOFIA cubelets (${\sim}54\%$ of our initial sample) in which the detected sources above the S/N thresholds are too small too attempt kinematic modeling. Specifically, they are either spatially less extended than the beam area or spectrally less extended than two channels.

We visually checked the output of the \bbarolo\ \texttt{SEARCH} task. In 217 of the remaining 2751 AWES cubelets, the \texttt{SEARCH} task mistakenly detected an individual object as two separate sources. The most common examples are \hi disks at moderate to high inclination and with central depressions, such that the low S/N in the inner regions leads to a separation of a galaxy into its approaching and receding side \citep[see also][]{2025A&A...703A.295Siljeg}. In such cases, we manually repeated the \texttt{SEARCH} task, where we first increased the channel separation for detection merging and, if this did not recover a single source, lowered the primary and secondary S/N thresholds. Similarly, 90 cubelets contain galaxies at one of the edges of the spectral range of the observation, so the \hi emission can be recovered for only a fraction of their spectral extent. We excluded these detections from our sample.

At low to moderate resolutions, kinematic modeling is particularly sensitive to the choice of initial geometric parameters. To determine reliable initial estimates for the centers of our galaxies, we crossmatched them with the Principal Galaxies Catalog \citep[PGC;][]{2003A&A...412...45Paturel}. If this did not result in a counterpart, we crossmatched them with the NASA/IPAC Extragalactic Database\footnote{\href{https://ned.ipac.caltech.edu/}{https://ned.ipac.caltech.edu/}} (NED), accepting counterparts that are classified as galaxies or listed in the Arecibo Legacy Fast ALFA (ALFALFA) Extragalactic \hi Source Catalog \citep{2018ApJ...861...49Haynes}. We used a separation threshold of \SI{1}{\arcmin} based on the \hi centers that were determined by the \bbarolo\ \texttt{SEARCH} task.

We obtained optical images for all of our galaxies, using the following priority scheme: (1)~$grz$ images from the Dark Energy Spectroscopic Instrument (DESI) Legacy Imaging Survey's data release~10 \citep{2019AJ....157..168Dey}; (2)~$grz$ images from the Sloan Digital Sky Survey's (SDSS's) data release~18 \citep{2023ApJS..267...44Almeida}; (3)~$grz$ images from the Panoramic Survey Telescope and Rapid Response System 1 (Pan-STARRS1) data release~2 \citep{2016arXiv161205560Chambers}; and (4)~single-band images from the Digitized Sky Survey \citep[DSS;][]{1990AJ.....99.2019Lasker, 2000ASPC..216..145McLean}. We performed a preliminary visual classification of the sample based on these images. In particular, we removed detections corresponding to small clouds or clumps around the main \hi disks of the galaxies. Next, we corrected erroneous counterparts and added additional manual counterparts where the automated crossmatching had missed them. These fixes affected 247 \bbarolo\ detections. Moreover, 119 galaxies are present in more than one cubelet, because the initial SOFIA source finding was performed on individual AWES fields, which can overlap. In such cases, we only kept the cubelet for which the respective galaxy is located closest to the cubelet center. We also flagged interacting galaxies, which we identified based on the presence of one or more interacting partners in the optical images or in the \hi moment maps. We then added additional PGC or NED counterparts for these partners. Some of the original SOFIA data cubelets can therefore correspond to multiple objects in our galaxy sample. In total, this left us with 1924 individual galaxies, 67 ($\sim3\%$) of which are flagged as interacting.

Of this sample, 1278 galaxies were assigned a PGC counterpart, 52 an ALFALFA counterpart, and 513 some other NED counterpart. The remaining 81 galaxies (${\sim}4\%$) ended up lacking a named counterpart in PGC or ALFALFA or a counterpart classified as a galaxy in NED. All of these sources except for five lie outside of the ALFALFA footprint. Upon visual inspection, 69 of these 81 galaxies have visible counterparts in the optical images, even though they are often faint and surrounded by many foreground stars. A further ten galaxies have extremely bright foreground stars close to the \hi coordinates, which makes the detection of potential counterparts impossible. The remaining two galaxies without counterparts are both located close to the Galactic plane and optical images for them are only available from DSS and Pan-STARRS.

\begin{figure*}[t]
    \centering
    \includegraphics[width = \textwidth]{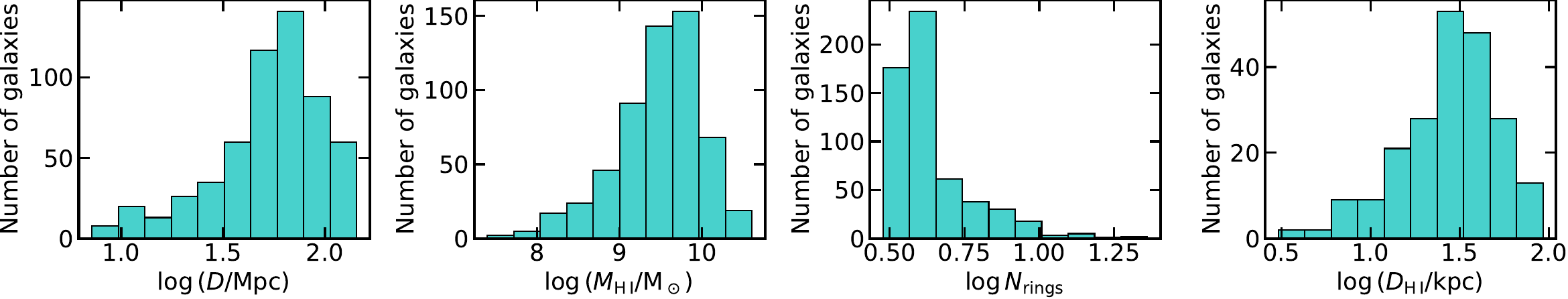}
    \caption{Selected properties of 568 AWES galaxies with reliable kinematic models. \textit{From left to right:} Galaxy distance, $D$, from the CF4 flow model, total \hicap\ mass, $M_\mathhi$, from Eq.~\ref{eq_hi_mass}, number of rings, $N_\mathrm{rings}$, in the kinematic models (see Sect.~\ref{sec_kin}), and \hicap\ diameter, $D_\mathhi$ (see Sect.~\ref{sec_dist}). The \hi diameter could be determined for 213 galaxies with sufficiently deep and well-resolved data.}
    \label{fig_sample_hist}
\end{figure*}

\subsection{Distances, H\,{\small I} masses, and H\,{\small I} radii}  \label{sec_dist}

We calculated distances from systemic velocities using the Cosmicflows-4 (CF4) flow model \citep{2023A&A...670L..15Courtois, 2024NatAs...8.1610Valade}, which reconstructs the linear 3D velocity field based on ${\sim}38\,000$ galaxy groups with redshift-independent distances \citep{2023ApJ...944...94Tully}. The flow model is accessible as part of the Extragalactic Distance Database\footnote{\href{https://edd.ifa.hawaii.edu/}{https://edd.ifa.hawaii.edu/}} \citep{2009AJ....138..323Tully}. Our final distances are based on the more robust systemic velocities, $V_\mathrm{sys}$, obtained from kinematic modeling (see Sect.~\ref{sec_kin}), while the initial distances for kinematic modeling are based on the central velocities, $V_\mathrm{cen}$, of the 20\% flux levels of the spatially integrated \hi spectra from the \bbarolo\ \texttt{SEARCH} task. The CF4 flow model is consistent with a Hubble constant of $H_0 = \SI{75}{\kilo\meter\per\second\per\mega\parsec}$ \citep{2023ApJ...944...94Tully}.

Distances are a dominant source of error in many physical properties of galaxies (e.g., masses and sizes), so realistic uncertainties should be propagated into these quantities. This issue was empirically explored by \citet{2025A&A...696A.185Haubner}, who compared the distances from four flow models, including CF4, to redshift-independent distances of 1956 galaxies \citep{2023ApJ...944...94Tully}. The resulting uncertainties depend only on the flow distance of a galaxy itself and on its angular separation from the Virgo cluster. For our AWES sample, this uncertainty scheme returns values between ${\sim}0.04$ and ${\sim}\SI{0.17}{\dex}$, which we propagate into \hi masses and diameters.

Flow models can have multiple solutions where one observed systemic velocity corresponds to several (three in the case of CF4) possible distances. This degeneracy results from strong peculiar motions in the vicinity of large mass concentrations such as galaxy clusters. In our initial sample of 1924 galaxies, 28 galaxies are affected by this. For the results of our kinematic modeling pipeline, the assumed distance is mostly negligible. Similarly, distance errors only move galaxies along the \hi mass-size relation because of its slope of roughly two. Therefore, we simply assumed the nearest of the CF4 distances for these analyses when there are multiple values available. However, for future investigations based on our rotation curves (i.e., detailed mass modeling or BTFR studies), it will be crucial to determine which flow solution allows for the best description of the data.

We calculate total \hi masses via \citep[Eq.~48]{2017PASA...34...52Meyer}
\begin{equation}    \label{eq_hi_mass}
    \frac{M_\mathhi}{\SI{}{\solarmass}} \simeq 49.7\left(\frac{D}{\SI{}{\mega\parsec}}\right)^2\left(\frac{S}{\SI{}{\jansky\hertz}}\right),
\end{equation}
with the luminosity distance, $D$, and the total integrated flux, $S$, of a galaxy. As the total fluxes derived within the \bbarolo\ masks are systematically lower than fluxes from the ALFALFA survey, we used circular apertures around the galaxies' centers to measure their \hi fluxes instead (see Appendix~\ref{app_flux} for details). These apertures have radii that are \SI{30}{\arcsec} larger than the extent of the \bbarolo\ models. We corrected the \hi fluxes within them for a remaining constant 15\% discrepancy with ALFALFA. We assumed a random flux error of 15\% from the flux scale calibration, primary beam correction, and mosaicking uncertainties of Apertif \citep{2022A&A...667A..39Kutkin}. This value is consistent with the statistical uncertainty that we estimated from a comparison of Apertif and ALFALFA fluxes (see Appendix~\ref{app_flux}).

Furthermore, we derived \hi diameters $D_\mathhi$ for studying the \hi mass-size relation. These are defined as the diameters where the azimuthally averaged and deprojected \hi surface density profiles reach \SI{1}{\solarmass\per\square\parsec}. Kinematic modeling with \bbarolo\ provides the necessary surface density profiles, which are measured from the average gas surface densities within the model rings. We multiplied these profiles by $(1 + z)^4$ to correct for cosmological surface brightness dimming \citep[Eq.~74]{2017PASA...34...52Meyer} and additionally corrected them for the 15\% missing flux compared to ALFALFA.  The cosmological corrections can reach the level of ${\sim}10\%$. We calculate $D_\mathhi$ from a linear fit to the data points that bracket a deprojected surface density of $\Sigma_\mathhi = \SI{1}{\solarmass\per\square\parsec}$. We corrected the resulting \hi diameters for beam smearing using the Gaussian approximation of \citet{2016MNRAS.460.2143Wang}, that is,
\begin{equation}
    D_\mathhi = \sqrt{{D_\mathhi^\mathrm{obs}}^2 - b_\mathrm{maj} b_\mathrm{min}},
\end{equation}
with $D_\mathhi^\mathrm{obs}$ as the uncorrected \hi diameter and $b_\mathrm{maj}$ and $b_\mathrm{min}$ the major and minor beam full width at half maximum in physical units, respectively. We do not provide \hi diameters for galaxies with $D_\mathhi^\mathrm{obs} < 2\sqrt{b_\mathrm{maj}b_\mathrm{min}}$, for which the correction is very uncertain.

\subsection{Sample properties}

Figure~\ref{fig_sample_hist} gives an overview of the properties of the 568 galaxies for which we were able to derive reliable kinematic models and which make up the body of our data release (see Sect.~\ref{sec_results}). The panels show the flow distance distribution, \hi mass distribution, distribution of the number of rings, $N_\mathrm{rings}$, in the kinematic models, and distribution of \hi diameters of the galaxies. Distances range from ${\sim}7$ to ${\sim}\SI{143}{\mega\parsec}$, with most galaxies between 30 and \SI{100}{\mega\parsec}. The median distance is ${\sim}\SI{60}{\mega\parsec}$. \hi masses span three orders of magnitude from ${\sim}\SI{2.5e7}{}$ to ${\sim}\SI{4e10}{\solarmass}$, with a median value of ${\sim}\SI{3.5e9}{\solarmass}$. Based on the relation between $M_\mathhi$ and stellar mass $M_\star$ \citep{2012ApJ...756..113Huang, 2015MNRAS.447.1610Maddox}, this implies that our sample contains substantial contributions from both massive galaxies ($M_\star > \SI{e9}{\solarmass}$) and dwarf galaxies ($M_\star < \SI{e9}{\solarmass}$).

The number of rings in the kinematic models ranges from 3 to 23, with a strong peak (234 galaxies) at 4 rings. These resolutions are mainly a result of the necessary smoothing of our data to obtain high enough S/N for kinematic modeling. Finally, the \hi diameters range from ${\sim}3$ to ${\sim}\SI{90}{\kilo\parsec}$. The median $D_\mathhi$ is ${\sim}\SI{30}{\kilo\parsec}$. The determination of the diameter was possible for 213 of the 568 galaxies with reliable kinematic models. Specifically, the observations are deep enough to reach a surface density of \SI{1}{\solarmass\per\square\parsec} for 228 galaxies. Of these, 15 galaxies were excluded because they are so poorly resolved that the beam smearing correction is very uncertain.

\section{Kinematic modeling}    \label{sec_kin}

\subsection{Standard pipeline}  \label{sec_std_pipeline}

The primary requirements for the kinematic modeling of a large \hi dataset such as that of AWES are
\begin{enumerate}
    \item[(1)] a high degree of automation at a low computational cost;
    \item[(2)] homogeneity to keep systematics under control; and
    \item[(3)] reliability also for poorly resolved galaxies with low S/N.
\end{enumerate}
Starting from these considerations, we based our kinematic pipeline on the 3D-based Analysis of Rotating Objects from Line Observations\footnote{\href{https://editeodoro.github.io/Bbarolo/}{https://editeodoro.github.io/Bbarolo/}} \citep[\bbarolo{};][]{2015MNRAS.451.3021DiTeodoro} code, namely, development version~1.7.8. Its \texttt{3DFIT} task allows for the construction of 3D tilted-ring models convolved with the spatial and spectral resolution of the respective observation. Consequently, it can account for instrumental effects, including beam smearing \citep[e.g.,][]{1978PhDT.......195Bosma, 1987PhDT.......199Begeman}, and recover the underlying rotation velocities also for galaxies with only few resolution elements \citep{2015MNRAS.451.3021DiTeodoro, 2025ApJ...994..232Deg}. Furthermore, the code is highly automatable and has routines for estimating initial model parameters on its own.

In the tilted-ring modeling approach, each ring is characterized by eight geometric and kinematic parameters; namely, the center (\texttt{XPOS}, \texttt{YPOS}), the systemic velocity, \texttt{VSYS}, the inclination, \texttt{INC}, the position angle, \texttt{PA}, the disk thickness, \texttt{Z0}, the rotation velocity, \texttt{VROT}, the velocity dispersion, \texttt{VDISP}, and the radial velocity, \texttt{VRAD}. \bbarolo\ sets the ring width to half a beam width if a galaxy is resolved with less than 20 beams along the semi-major axis and to exactly one beam width otherwise. All of our galaxies belong to the former case. We fixed the vertical profile of the model disks to Gaussian functions with a scale height, \texttt{Z0}, of \SI{100}{\parsec}. This assumption has a negligible effect as the intrinsic vertical structure of the disks is practically unconstrained at our physical resolutions ($>\SI{1}{\kilo\parsec}$). Additionally, we fixed all radial velocities to zero. The spectral and spatial resolutions of our data are insufficient to measure them as they are usually subdominant compared to the rotation velocity and degenerate with warps in the position angle \citep{2012A&A...544A.145Lelli, 2021ApJ...923..220DiTeodoro}. The other six parameters were fit.

\begin{table}[b]
    \caption{\bbarolo\ parameters used in our kinematic pipeline.}
    \label{table_parameters}
    \centering
    \begin{tabular}{ccc}
        \hline\hline
        Parameter & Value & Comment \\
        \hline
        \textbf{Global} & &\\
        \texttt{LTYPE} & 1 & Gaussian vertical profile\\
        \texttt{TWOSTAGE} & True & \\
        \texttt{REGTYPE} & \texttt{AUTO} & Parameter regularization\\
        \texttt{NORM} & \texttt{AZIM LOCAL} & Model $\Sigma_\mathhi$ normalization\\
        \texttt{DELTAINC} & \SI{20}{\deg} & Search range around \texttt{INC}$^0$\\
        \texttt{DELTAPA} & \SI{20}{\deg} & Search range around \texttt{PA}$^0$\\
        \texttt{CDENS} & 100 & Cloud surface density\\
        \texttt{LINEAR} & 0.3 & See Appendix~\ref{app_lsf}\\
        \texttt{VELDEF} & \texttt{RELATIVISTIC} & \\
        \hline
        \textbf{Specific} & &\\
        \texttt{XPOS}$^0$ & Crossmatching & See Sect.~\ref{sec_prep}\\
        \texttt{YPOS}$^0$ & Crossmatching & See Sect.~\ref{sec_prep}\\
        \texttt{VSYS}$^0$ & 0 & See \texttt{REDSHIFT}\\
        \texttt{REDSHIFT} & $V_\mathrm{cen}/c$ & From \texttt{SEARCH}\\
        \texttt{INC}$^0$ & Automated & \\
        \texttt{PA}$^0$ & Automated & \\
        \texttt{VROT}$^0$ & $W_{20}/2$ & From \texttt{SEARCH}\\
        \texttt{VDISP}$^0$ & \SI{8}{\kilo\meter\per\second} & \\
        \texttt{Z0} & $\SI{100}{\parsec}/D$\,(arcsec) & See Sect.~\ref{sec_dist} for $D$\\
        \hline
    \end{tabular}
    \tablefoot{Parameters with the superscript 0 were allowed to vary during the fits. For details on individual parameters, see Sect.~\ref{sec_std_pipeline}. Parameters that are not listed were kept at their standard values.}
\end{table}

To avoid unphysical discontinuities in the geometric parameters, we used \bbarolo{}'s \texttt{TWOSTAGE} mode, where two subsequent kinematic fits are performed. In the first one, all parameters are free in each ring, while in the second one, only \texttt{VROT} and \texttt{VDISP} are free. During this second fit, (\texttt{XPOS}, \texttt{YPOS}) and \texttt{VSYS} are set to their median values across all rings. To account for possible disk warps, \texttt{INC} and \texttt{PA} are regularized by using their median value, a polynomial function, or a Bézier curve. We set the \texttt{REGTYPE = AUTO} flag, so the exact choice of regularization is determined by \bbarolo\ depending on the best-fit parameter values of the first kinematic model.

\bbarolo\ does not fit for the surface density profiles, but it normalizes the surface density of the tilted-ring models to the observed gas distribution of a galaxy. This can be done on a ring-by-ring basis using the azimuthally averaged \hi moment-zero map (\texttt{NORM = AZIM}) or on a pixel-by-pixel basis using the observed \hi moment-zero map (\texttt{NORM = LOCAL}). The local normalization has the advantage that it can produce reliable kinematic models for galaxies that have regular kinematics, but irregular morphologies. Its main disadvantage is that it can obscure wrong inclination estimates in the model because the flux contours of the data are reproduced by construction. Therefore, to combine the advantages of both methods, we performed the first fit of the \texttt{TWOSTAGE} routine with the azimuthal normalization and the second fit with the local normalization. This allows us to identify models with wrong inclination estimates from the channel maps of the azimuthally averaged models, while retaining the more robust local normalizations for our final kinematic models.

The success or failure of the tilted-ring modeling procedure depends on the choice of initial parameters, in particular, initial geometric parameters. In Table~\ref{table_parameters} we give an overview of our choice of parameters, separated into global ones that are the same for every galaxy and specific ones that vary from galaxy to galaxy. We allowed \texttt{INC} and \texttt{PA} to vary in a range of $\pm\SI{20}{\deg}$ around their initial estimates (see below). Furthermore, we set \texttt{CDENS = 100}, that is, the number of model \hi clouds per spaxel in the plane of the model galaxy in units \SI{e20}{\per\square\centi\meter} (prior to the surface density normalization). We increased this value compared to its standard setting (10) to allow for a better sampling of the relatively low-S/N Apertif galaxies.

\bbarolo\ assumes Gaussian spectral responses during kinematic modeling to take the instrumental line broadening in the data cubelets into account. This instrumental broadening is set through the \texttt{LINEAR} parameter, which is the standard deviation of the observation's line spread function in units of channels. However, the spectral response functions of the Apertif cubelets are boxcars with a width of one channel (\SI{7.7}{\kilo\meter\per\second}). The spectral resolution is fortunately still good enough compared to the intrinsic velocity dispersion of galactic \hi to assume a Gaussian line response. We show this explicitely in Appendix~\ref{app_lsf}, where we also motivate our choice of \texttt{LINEAR = 0.3}.

\bbarolo\ allows for the automated estimation of initial parameters for kinematic modeling \citep[for details see][]{2015MNRAS.451.3021DiTeodoro}. We used this option for the inclination, \texttt{INC}, and the position angle, \texttt{PA}. With respect to the coordinates of a galaxy, we took the initial estimate of the kinematic center (\texttt{XPOS} and \texttt{YPOS}) from the PGC or NED counterpart of each galaxy. For galaxies without PGC or NED counterparts, we used the estimates of the geometric \hi centers provided by the \texttt{SEARCH} task. The initial estimate for \texttt{VSYS} was set to zero because we perform kinematic modeling in the rest frames of the galaxies by setting the redshift parameter $\texttt{REDSHIFT} = V_\mathrm{cen}/c$, where $V_\mathrm{cen}$ was taken from the global \hi line profile produced by the \texttt{SEARCH} task. The redshift remains fixed during kinematic modeling, while adjustments to the systemic velocity are made through the \texttt{VSYS} parameter. The deredshifting also has the effect that the precise choice of the velocity convention (\texttt{VELDEF}) has a negligible effect. Next, we set the initial rotation velocity \texttt{VROT} to $W_{20}/2$, where $W_{20}$ is the width of the global \hi line profiles at 20\% of the peak flux as measured by the \texttt{SEARCH} task. For the initial velocity dispersion, \texttt{VDISP}, we used a constant value of \SI{8}{\kilo\meter\per\second} for all galaxies, in line with high-resolution observations of nearby galaxies \citep[e.g.,][]{2009AJ....137.4424Tamburro}.

\subsection{Edge-on galaxies}

The rotation curves of highly inclined galaxies are barely affected by uncertainties in the inclination estimates. However, fitting approaches that work on a ring-by-ring basis become unreliable above inclinations of ${\sim}\SI{80}{\deg}$ because the assumption that any line of sight to the observer crosses the galaxy at only one galactocentric radius is violated. The line profiles in the inner parts of such edge-on galaxies are asymmetrically broadened because of the alignment of \hi emission at different galactocentric radii and with different projected radial velocity components. Attempts to model these line profiles by fitting each ring separately typically result in underestimated rotation velocities and overestimated velocity dispersions in the inner parts of such galaxies, though the problem can be overcome in 1D by tracing the envelope of the position-velocity diagram \citep[PVD;][]{1979A&A....74...73Sancisi} and in 3D by using software that fits a tilted-ring model globally \citep[e.g., KinMS or FAT;][]{2013MNRAS.429..534Davis, 2015MNRAS.452.3139Kamphuis}.

For this reason, we excluded galaxies with an optical inclination larger than \SI{80}{\deg} from the standard kinematic modeling pipeline and instead fit them with a dedicated edge-on pipeline. We calculated galaxy inclinations via \citep{1926ApJ....64..321Hubble}
\begin{equation}
    \cos^2{i} = \frac{(b/a)^2 - q_0^2}{1 - q_0^2},
\end{equation}
where we used the optical axis ratio $b/a$ from the PGC catalog and an intrinsic axis ratio of $q_0 = 0.2$ \citep{1990ApJ...349....1Fouque}. Edge-on galaxies that were still fit with the standard pipeline because of a wrong or missing PGC inclination were flagged during the visual inspection of the models (see Sect.~\ref{sec_insp}).

For the edge-on pipeline, we fixed the kinematic model's inclination \texttt{INC} to the value calculated from the PGC axis ratio\footnote{We used $i = \SI{89}{\deg}$ for $b/a \leq q_0$, as an inclination of \SI{90}{\deg} causes the \bbarolo\ kinematic fits to fail due to numerical effects.} to counteract \bbarolo{}'s frequent underestimation of the inclination of edge-on galaxies. Furthermore, we fixed the velocity dispersion \texttt{VDISP} to a constant value of \SI{8}{\kilo\meter\per\second}. The velocity dispersions of edge-on galaxies are poorly constrained as the observed line broadening is dominated by projection effects. We tested and confirmed that the choice of fixing \texttt{VDISP} has a negligible impact on the other parameters of the final kinematic models.

The main difference to the standard pipeline is that the two fits of the \texttt{TWOSTAGE} mode (with fixed \texttt{INC} and \texttt{VDISP} for edge-on galaxies) are followed up with a third fit that uses the \texttt{REVERSE = True} flag. In the reverse mode, \bbarolo\ starts the kinematic modeling procedure at the outermost ring and moves inward, as opposed to the outward modeling direction of the regular fits. This allows \bbarolo\ to obtain a good model of the outermost tilted ring, which is less affected by projections of other galactocentric radii. Subsequently, the code can account for intervening gas from larger galactocentric radii while moving further inward \citep[see also][]{2004MNRAS.352..787Kregel}. However, the reverse mode is even more sensitive to the choice of initial geometric parameters than the standard fitting mode, which is why (\texttt{XPOS}, \texttt{YPOS}), \texttt{VSYS}, and \texttt{PA} are determined through the initial \texttt{TWOSTAGE} fits. The only free parameter in the reverse fits is \texttt{VROT}.

\subsection{Correction for pressure support}    \label{sec_adc}

We corrected the rotation velocities, $V_\mathrm{rot}$, of our kinematic models for pressure support to estimate the circular velocities, $V_\mathrm{circ}$, of test particles in the galactic gravitational potentials. This correction is typically negligible for \hi disks, but can become important for dwarf galaxies with $V_\mathrm{rot} \lesssim \SI{50}{\kilo\meter\per\second}$ \citep[e.g.,][]{1996AJ....111.1551Meurer, 2014A&A...566A..71Lelli, 2017MNRAS.466.4159Iorio}. For this purpose, we used the asymmetric drift correction $V_\mathrm{ADC}^2$, which allows us to write the circular velocity as
\begin{equation}
    V_\mathrm{circ}^2 = V_\mathrm{rot}^2 + V_\mathrm{ADC}^2.
\end{equation}
The asymmetric drift correction itself can be calculated as \citep[Eq.~4.228]{2008gady.book.....Binney}
\begin{equation}    \label{eq_adc}
    V_\mathrm{ADC}^2 = \sigma_V^2\left(-\frac{\partial \ln{\Sigma_\mathhi}}{\partial \ln R} - \frac{\partial \ln{(\sigma_V^2)}}{\partial \ln R}\right),
\end{equation}
where we assume that the velocity dispersion $\sigma_V$ is isotropic and independent of the height above the galactic plane and that the vertical gas density profile is constant with radius.

To allow for an easy analytical differentiation of $\sigma_V(R)$ and $\Sigma_\mathhi(R)$, we regularized them with polynomial functions. The surface density profile is robustly measured from azimuthal averages of the moment-zero map, while the velocity dispersion profile $\sigma_V(R)$ is kinematically fit during the modeling. Therefore, we do not use the same polynomial degrees for both profiles, but allow for more complicated functions when regularizing $\Sigma_\mathhi(R)$. After testing a range of schemes, we decided to determine the polynomial degree $d$ for fitting $\Sigma_\mathhi(R)$ through
\begin{equation}    \label{eq_regularize}
    d = \mathrm{int}\left[\frac{40}{\pi}\arctan{\left(\frac{N_\mathrm{rings}}{16}\right)}\right],
\end{equation}
where the arctangent function is evaluated in radians and int indicates the removal of decimal places. This function has an asymptotic value of 19, which sets the maximum polynomial degree that is allowed for $\Sigma_\mathhi(R)$. The actual value that is reached by the AWES kinematic model with the most rings (22) is eleven. At the low end, the polynomial degree is three for $N_\mathrm{rings} = 4$. For $N_\mathrm{rings} = 3$, we regularized $\Sigma_\mathhi(R)$ with a linear fit. Models with only two rings are rejected during visual inspection.

In some cases, the polynomial fits can still result in unphysically large values of $V_\mathrm{ADC}$, when a small variation in $\Sigma_\mathhi(R)$ at the outer edges of a kinematic model prompts an inflection point in the regularization functions. To suppress this behavior, only while fitting the polynomials, we added a single "phantom data point" outside of the final data point $(R_\mathrm{out}, \Sigma_\mathhi^\mathrm{out})$ of each $\Sigma_\mathhi(R)$ profile. This additional data point has the value $(R_\mathrm{out} + R_\mathrm{ring}, \Sigma_\mathhi^\mathrm{out})$, where $R_\mathrm{ring}$ is the width of one ring, and suppresses strong variations in the surface density at the edges of the models.

\begin{figure}[t]
    \centering
    \includegraphics[width = \columnwidth]{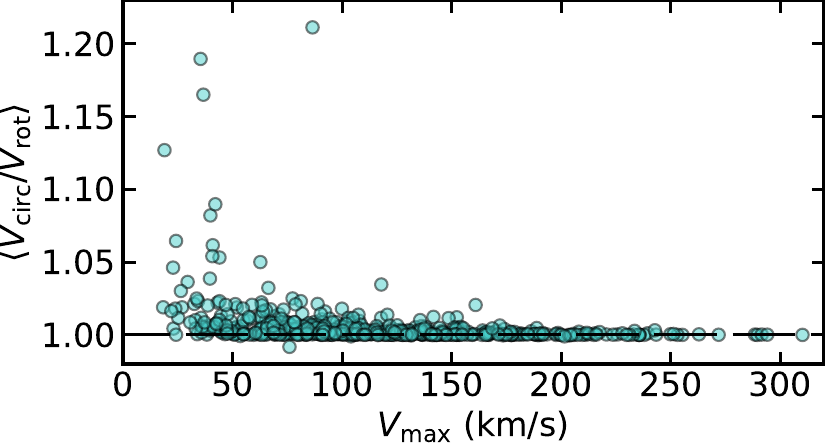}
    \caption{Mean value of $V_\mathrm{circ}/V_\mathrm{rot}$ against the maximum circular rotation velocity $V_\mathrm{max}$ for each of the 568 galaxies with reliable kinematic models. The dashed horizontal line marks a ratio of one.}
    \label{fig_adc}
\end{figure}

For the velocity dispersion profile $\sigma_V(R)$, we used a more conservative regularization scheme. These regularized values enter the asymmetric drift correction not only through the derivative of $\sigma_V$, but also through its absolute value:
\begin{description}
    \item[$N_\mathrm{rings}\leq 5$:] We used the median value of $\sigma_V(R)$, which means that the derivative of $\sigma_V(R)$ in Eq.~\ref{eq_regularize} vanishes. The value of $\sigma_V(R)$ still determines the overall normalization of the ADC.
    \item[$N_\mathrm{rings} > 5$:] We fit $\sigma_V(R)$ with a polynomial of degree three.
\end{description}
The overall importance of the ADC for our kinematic models is illustrated in Fig.~\ref{fig_adc}. It shows the mean value of $V_\mathrm{circ}/V_\mathrm{rot}$ for each of the 568 galaxies with an acceptable kinematic model plotted against the maximum circular velocity $V_\mathrm{max}$. The correction tends to become more important for lower $V_\mathrm{max}$, but the vast majority of galaxies has $V_\mathrm{circ}/V_\mathrm{rot} \simeq 1$. There are only 12 galaxies where the correction makes a difference larger than 5\%, all of which have $V_\mathrm{max} < \SI{100}{\kilo\meter\per\second}$. This implies that, as expected, the rotation velocity is a reliable tracer of the underlying gravitational potential for most \hi disks.

An illustration of the derivation of the asymmetric drift correction for an individual galaxy can be seen in Fig.~\ref{fig_a1_plot1}. This figure shows the overview of a reliable kinematic model, as described in Sect.~\ref{sec_insp}. The solid blue lines in the surface density and velocity dispersion panels show the polynomial fits to the profiles of these parameters (degree five for $\Sigma_\mathhi(R)$ and degree three for $\sigma_V(R)$). The actual values of $V_\mathrm{ADC}$ are depicted with the solid blue line in the rotation velocity plot, while the resulting circular velocity is given by the dashed blue line. Below roughly \SI{30}{\arcsec}, the correction is negative and $V_\mathrm{ADC}$ itself is formally imaginary because of the positive slopes of the surface density and velocity dispersion profiles.

\subsection{Uncertainties}  \label{sec_unc}

\bbarolo\ allows for the estimation of  parameter uncertainties through Monte Carlo sampling of parameters around their best-fit values. This procedure is computationally expensive and does not account for covariances as parameters are varied individually. More importantly, Monte Carlo estimates of the formal uncertainty in $V_\mathrm{rot}$ may not capture the real physical uncertainties due to non-circular motions, asymmetries, or other failures of the assumptions of tilted-ring modeling. Consequently, the advantage of the method is limited, particularly with respect to the necessity of performing such estimations for almost two thousand kinematic models. 

Therefore, for the rotation velocities, $V_\mathrm{rot}$, and the velocity dispersions, $\sigma_V$, we estimate the uncertainty from the difference between the approaching and the receding side of the disk \citep[e.g.,][]{1999PhDT........27Swaters, 2008AJ....136.2648DeBlok}. We performed two additional kinematic fits to each galaxy, where we set the \texttt{SIDE} parameter to \texttt{A} and \texttt{R} for considering only the approaching and the receding side, respectively. In these fits, all geometric parameters were fixed to the final best-fit values from the two-sided fitting (or the reverse fit for edge-on galaxies) and only $V_\mathrm{rot}$ and $\sigma_V$ were allowed to vary. Assuming that the difference between the approaching and the receding side estimates the $1\sigma$ interval around $V_\mathrm{rot}$, the uncertainty follows as
\begin{equation}
    \delta V_\mathrm{rot} = \frac{|V_\mathrm{app} - V_\mathrm{rec}|}{2},
\end{equation}
with $V_\mathrm{app}$ the rotation velocity of the approaching side and $V_\mathrm{rec}$ the rotation velocity of the receding side.

This method can capture deviations from a tilted-ring geometry, although it provides only a first estimate of the true $1\sigma$ uncertainties. In particular, it might be an underestimate in the marginally resolved regime, where the resulting uncertainties are sometimes very small. Therefore, we assign $\delta V_\mathrm{rot} = \Delta V/(2\sqrt{2\ln{2}})$ as a lower limit, where $\Delta V$ is the channel width. Furthermore, a fit to the approaching or receding side can sometimes fail, either completely or only in specific rings. For example, this can happen for asymmetric gas distributions where one half of a galaxy does not provide sufficient constraints for kinematic modeling at a specific radius. In these cases, we assign a conservative 25\% uncertainty ($\delta V_\mathrm{rot} = V_\mathrm{rot}/4$), to reflect the potentially disturbed state of the gas distribution. We assign uncertainties on $\sigma_V$ in the same way as those on $V_\mathrm{rot}$. Neglecting the uncertainties from the radial derivatives in Eq.~\ref{eq_adc}, the final uncertainty on $V_\mathrm{circ}$ is given by 
\begin{equation}
    \delta V_\mathrm{circ} = \sqrt{\left(\frac{V_\mathrm{rot}}{V_\mathrm{circ}}\delta V_\mathrm{rot}\right)^2 + \left(\frac{\sigma_V}{V_\mathrm{circ}}\delta \sigma_V\right)^2}.
\end{equation}

Given that $V_\mathrm{rot}$ varies with the galaxy inclination as $\sin^{-1}{i}$, its final uncertainty also depends on the inclination uncertainty $\delta i$. However, an error in inclination changes only the overall normalization of the rotation curve of a flat disk. Therefore, $\delta i$ acts as a systematic error, which should not be included in $\delta V_\mathrm{rot}$, but instead be appropriately considered depending on the specific analysis that is being performed. Typical values of $\delta i$ are ${\sim}\SI{5}{\deg}$ \citep[e.g.,][]{2016AJ....152..157Lelli, 2022PASA...39...59Deg}, although smaller values may be appropriate for edge-on galaxies and larger values (${\sim}\SI{10}{\deg}$) for face-on galaxies. The reason for this is the difficulty of measuring $i$ for mildly inclined disks, as the axis ratio is proportional to $\cos{i}$, combined with the intrinsic deviation of galactic disks from perfectly circular shapes \citep[e.g.,][]{2020ApJ...900..163KadoFong, 2024A&A...689L...3Lelli}.

The inclination uncertainties affect the rotation velocity $V_\mathrm{rot}$ in a substantial way at lower inclinations and typically become dominant below \SI{40}{\deg}. Therefore, such galaxies should be excluded from analyses that depend on the actual amplitude of the rotation curves, such as studying the BTFR and the RAR. However, the rotation curve shapes do not suffer from the same problem and can still be used, for example, to perform fits in which the inclination is treated as a nuisance parameter \citep[e.g.,][]{2018A&A...615A...3Li, 2020ApJS..247...31L}. Similar considerations also apply to the deprojected \hi surface density, which is calculated from the measured one by multiplying it by $\cos{i}$. In this case, inclination uncertainties affect edge-on galaxies most strongly and are negligible for face-on galaxies.

\subsection{Visual inspection and acceptance flags}    \label{sec_insp}

\begin{figure*}[p]
    \centering
    \includegraphics[width = \textheight - 40pt, angle = 90]{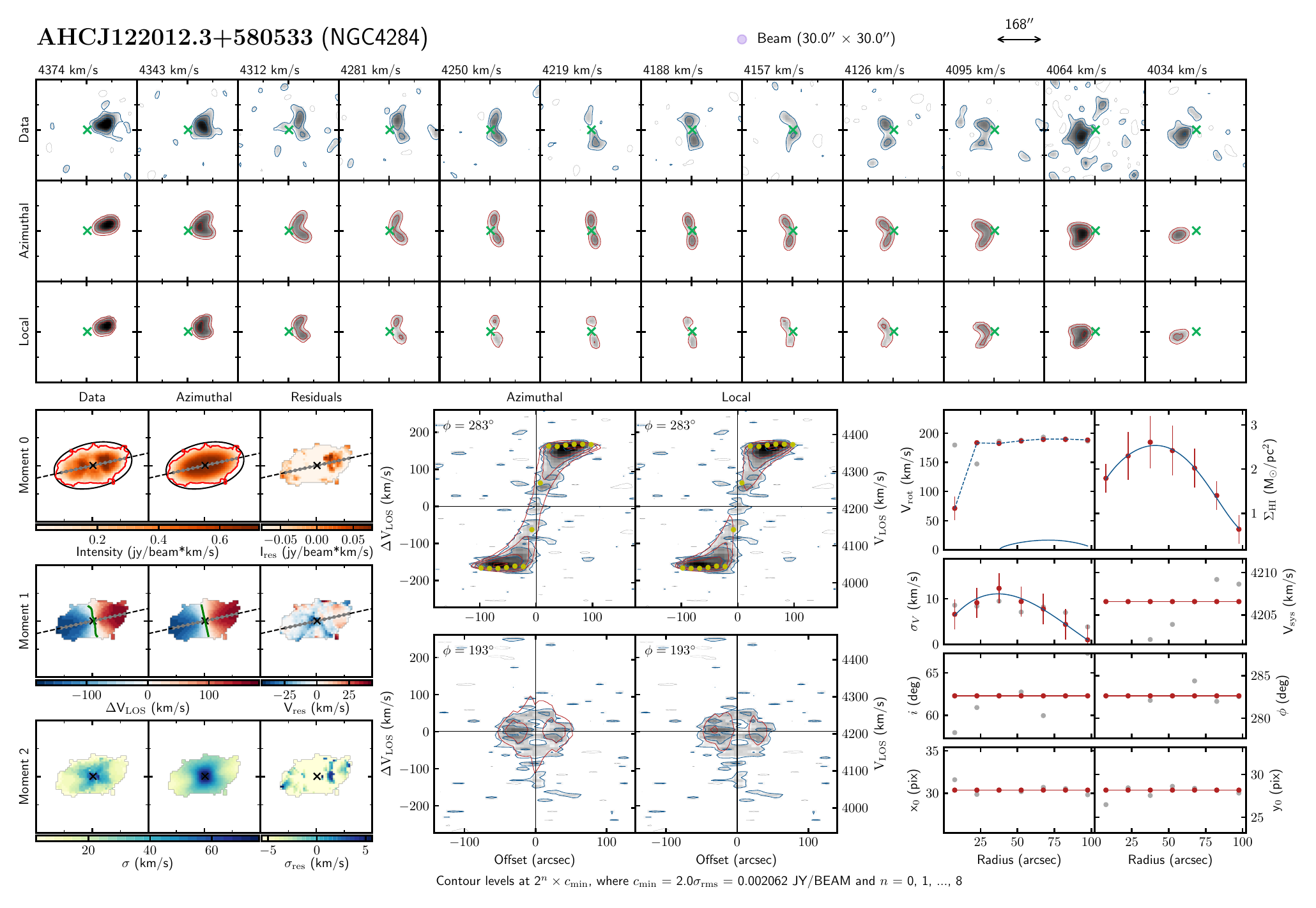}
    \caption{Example of the \bbarolo\ output for an $A = 1$ kinematic model, meaning that the model is acceptable and cannot be reasonably improved. The different panels show the channel maps of the data, the azimuthally normalized model, and the localy normalized model (\textit{left}), the moment maps of the data and the azimuthally normalized model (\textit{bottom-right}), the PVDs of both models for the major and the minor axis (\textit{center-right}), and the final parameters (\textit{top-right}). Blue lines in the panels of $V_\mathrm{rot}$, $\Sigma_\mathhi$, and $\sigma_V$ relate to the asymmetric drift correction described in Sect.~\ref{sec_adc}.}
    \label{fig_a1_plot1}
\end{figure*}

Kinematic modeling with \bbarolo\ can result in a formal success, even if the best-fit kinematic model provides an inadequate description of the data. This can happen when the code runs into local optima of the model parameter space. To date, there is no reliable method for flagging such inadequate kinematic models in an automated way. Therefore, we visually inspected each AWES \hi kinematic model and flagged it based on comparisons between the observed and the modeled channel maps, moment maps, and PVDs, which we also compared to the optical images of the galaxy. We assigned four values of the acceptance flag, $A$:
\begin{description}
    \item[$A = 1$:] Acceptable models that need not be improved further;
    \item[$A = 2$:] Models that are acceptable, but can still be improved;
    \item[$A = 3$:] Models that are not acceptable but can be modified to be;
    \item[$A = 4$:] Data that do not allow for a reliable kinematic model. 
\end{description}
$A = 1$ models provide excellent descriptions of the data, where the geometry of the galaxy and the shape and normalization of the rotation curve are correctly modeled. Figure~\ref{fig_a1_plot1} shows an overview of our assessment plots for such a model. $A = 2$ models also have reliable rotation curve shapes, but can have larger errors ($\lesssim 20\%$) in the rotation curve's overall amplitude, mostly from somewhat underestimated inclination angles or other minor errors in the modeled geometry. They are best used in conjunction with independent inclination estimates. $A = 1$ and $A = 2$ models comprise our data release. Examples of all four types of kinematic models are provided in Appendix~\ref{app_plots}.

We applied $A = 3$ to galaxies for which a reliable model can in principle be derived, but where the automated pipeline failed due to a wrong estimation of one or several parameters. Examples are kinematic centers, position angles, and inclinations that are wrong to such a degree that the shape of the rotation curve is incorrect. This includes edge-on galaxies that were not modeled with the edge-on pipeline. In contrast, $A = 4$ was assigned to galaxies that are strongly asymmetric, interacting, or otherwise disturbed to a degree that breaks the basic assumptions of tilted-ring modeling. This also applies to galaxies with insufficient S/N to allow for a reliable model assessment, and to galaxies with less than three rings in their rotation curves. The assignment of this flag is mostly related to properties of the data and of the galaxies, and not to genuine failures of \bbarolo\ itself.

Finally, some $A = 1$ and $A = 2$ models had velocity estimates at specific radii that are inconsistent with the data, while most measurements at other radii are reliable. The typical example is a generally good kinematic model with a sudden drop of ${\sim}\SI{100}{\kilo\meter\per\second}$ in the rotation velocity of the first or final ring. This can occur due to inadequacies in the \bbarolo\ masks, such as low S/N regions in the inner or outer \hi distributions that result in underestimated measurements of $V_\mathrm{rot}$. In such cases, we removed the individual affected velocity measurements and provide the remaining reliable kinematic model. This procedure does not affect the surface density profiles, which are measured from the moment-zero maps of the data and are generally more reliable. Nevertheless, the removal of velocity measurements in the kinematic models can lead to the rejection of a galaxy from our final sample if the number of rotation curve data points drops below three (see Sect.~\ref{sec_unresolved}).

\subsection{Poorly resolved galaxies}   \label{sec_unresolved}

Untargeted surveys, such as AWES and WALLABY, provide homogeneous and resolved \hi data for unprecedented numbers of galaxies, which allow for statistically powerful studies of galaxy evolution. However, the derived rotation curves do typically have lower resolutions than those of smaller, high-quality samples based on targeted observations \citep[e.g.,][]{2008AJ....136.2648DeBlok, 2016AJ....152..157Lelli}. Although our sample contains 97 rotation curves with at least six rings, the majority of our kinematic models have four rings. As an illustration, the minimum number of rotation curve rings for fitting a two-parameter DM halo and the three typical nuisance parameters (inclination, distance, and stellar mass-to-light ratio) is six. A larger number is desirable for constraining the exact shape of the rotation curve, measuring its correlation with the baryonic mass distribution \citep[e.g.,][]{2004IAUS..220..233Sancisi, 2012MNRAS.425.2299Swaters, 2016PhRvL.117t1101McGaugh}, and testing different DM halo profiles. In particular, this affects investigations of the long-standing core-cusp problem \citep{2010AdAst2010E...5DeBlok}, which can be studied best with high-resolution kinematic data of the centers of galaxies \citep[e.g.,][]{2008AJ....136.2648DeBlok, 2017MNRAS.466.1648Katz}. Similarly, \citet{2022PASA...39...59Deg} demonstrate that galaxies with less than eight beams across their major axis (corresponding to eight rings for our sampling) can have underestimated inner \hi surface density profiles, although their \hi diameters are more robust to resolution effects.

In general, the AWES rotation curves and surface density profiles should therefore be used with care when performing, for example, detailed mass modeling. Nevertheless, even four rings in a rotation curve can still allow for inferring global dynamical properties of galaxies \citep[e.g., halo masses;][]{2026MNRAS.548ag531Ponomareva}. A large untargeted sample, such as WALLABY or AWES, then acts as an important complement to higher resolution rotation curve samples because it allows for (1) more powerful statistical studies (e.g., of environmental effects or the resolved galaxy velocity function) and (2) a better understanding of selection effects, as the galaxies were selected blindly based on \hi flux.

The most poorly resolved rotation curves in our data release have three rings, which can still allow for the determination of the flat rotation velocity $V_\mathrm{f}$ \citep[e.g.,][]{2016ApJ...816L..14Lelli}. Since $V_\mathrm{f}$ is the measure of rotation velocity that minimizes the observed scatter in the BTFR \citep{2001ApJ...563..694Verheijen, 2019MNRAS.484.3267Lelli}, it is a physically significant characteristic of the kinematics of a galaxy. Such kinematic models can therefore offer an advantage over the use of integrated kinematic measures such as linewidths. This will however require careful assessments of the trustworthiness of a flat rotation curve with three rings.

Finally, the lowest number of rings in the kinematic models provided by \bbarolo\ is two. Overall, 361 of the 1924 models fall into this category. We assigned $A = 4$ to these models in most cases because they do not offer a significant advantage compared to global kinematic measurements. The exception are 150 of these galaxies with $A = 3$, for which a manual correction of the geometric parameters might in principle increase the number of rings to three or more due to a wrong estimation of the position angle.

\section{Results}   \label{sec_results}

\begin{figure*}[t]
    \centering
    \includegraphics[width = \textwidth - 60pt]{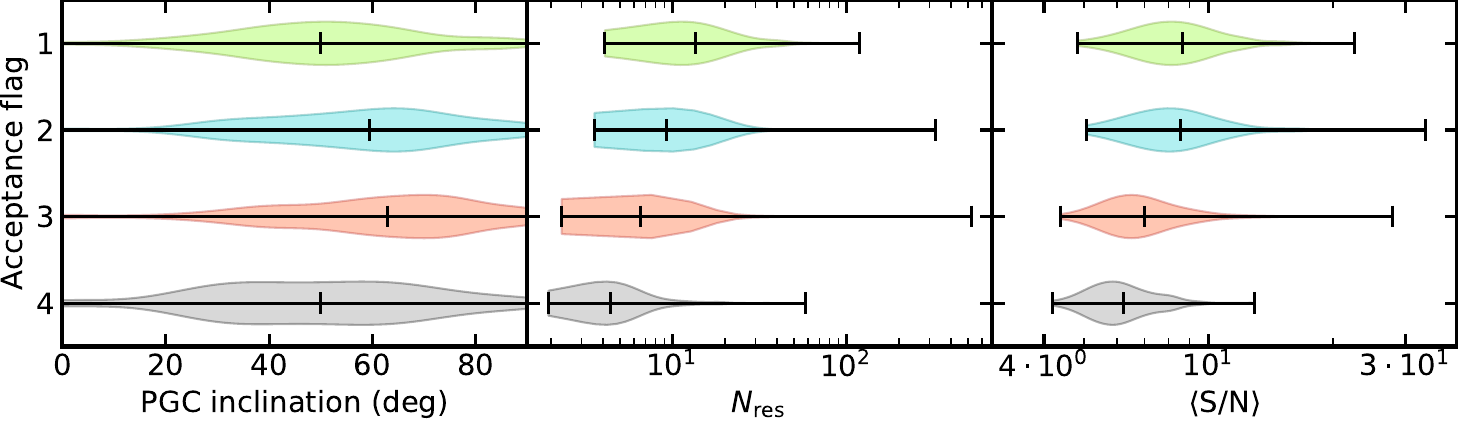}
    \caption{Violin plots of the distributions in PGC inclination, the number of resolution elements, $N_\mathrm{res}$, and the mean S/N within the \bbarolo\ masks for galaxies with different acceptance flags. The horizontal black lines indicate the full ranges of the galaxies in the respective subsamples and the central vertical black lines are the median values.}
    \label{fig_acceptance}
\end{figure*}

\subsection{Kinematic models}

The distribution of acceptance flags of the kinematic models is given in Table~\ref{table_samples}. Among the 1924 initial galaxies, 278 have $A = 1$, 290 have $A = 2$, 914 have $A = 3$, and 335 have $A = 4$. For the remaining 107 galaxies, the construction of the kinematic models failed, so we could not assess them visually. These galaxies are mostly in the extremely low S/N and extremely poorly resolved regime, such that \bbarolo\ resolved them with only two or three rings and failed to achieve convergence when fitting them. Overall, this leaves us with 568 galaxies with acceptable kinematic models, or a success rate of ${\sim}30\%$ after the S/N cut (see Sect.~\ref{sec_prep}). These galaxies form our data release.

\begin{table}[h]
    \caption{Sizes of the cube samples at different stages of the study, including the numbers of kinematic models that were assigned the four acceptance flags.}
    \label{table_samples}
    \centering
    \begin{tabular}{ll}
        \hline\hline
        Sample/subset & Number\\
        \hline
        Initial SOFIA cubelets & 5954\\
        Cubelets after S/N cut & 1924\\
        &\\
        \textbf{Kinematic modeling results} &\\
        $A = 1$ & 278 (14\%)\\
        $A = 2$ & 290 (15\%)\\
        \hspace{.3cm}$i$ too low & \hspace{.3cm}142 (49\%)\\
        $A = 3$ & 914 (48\%)\\
        \hspace{.3cm}Only two rings & \hspace{.3cm}150 (16\%)\\
        \hspace{.3cm}Wrong pipeline & \hspace{.3cm}132 (14\%)\\
        \hspace{.3cm}$i$ too low & \hspace{.3cm}115 (13\%)\\
        \hspace{.3cm}Wrong center & \hspace{.3cm}76 (8\%)\\
        \hspace{.3cm}$i$ too high & \hspace{.3cm}29 (3\%)\\
        $A = 4$ & 335 (17\%)\\
        \hspace{.3cm}Only two rings & \hspace{.3cm}208 (62\%)\\
        No output written & 107 (5\%)\\
        \hline
    \end{tabular}
    \tablefoot{For details on the acceptance flags, $A$, see Sect.~\ref{sec_insp}. Percentages of the overall acceptance flag numbers refer to the 1924 galaxies used for modeling. Percentages of the subsamples of $A = 2$, 3, and 4 models refer to the 290, 914, and 335 galaxies in these categories, respectively.}
\end{table}

\begin{figure*}[p]
    \centering
    \includegraphics[width = \textwidth - 20pt]{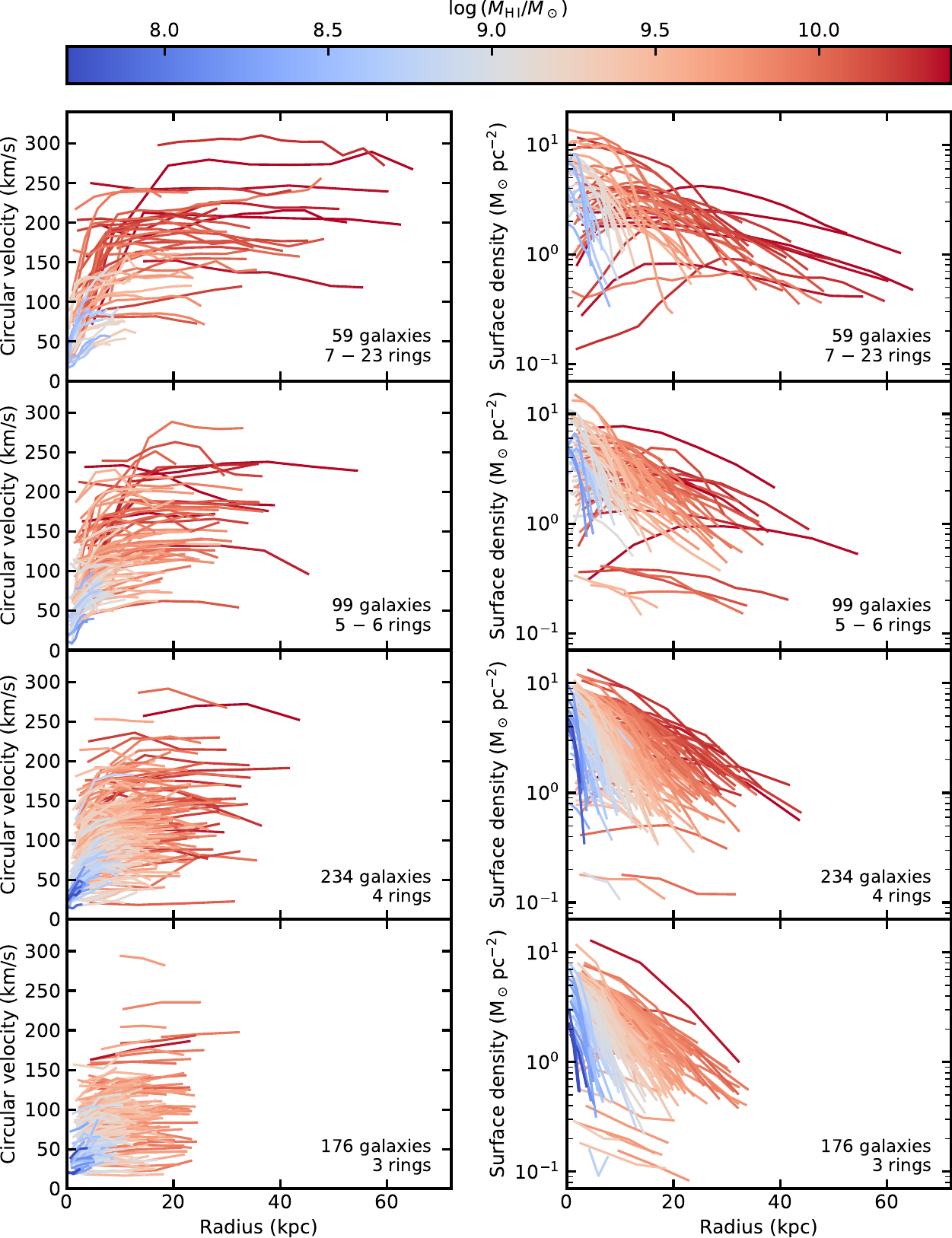}
    \caption{\hicap\ rotation curves and \hicap\ surface density profiles of all 568 galaxies with reliable kinematic models ($A = 1$ and $A = 2$). From top to bottom, the four panels depict kinematic models with a decreasing number of rings, as written in the bottom right corners. The rotation curves and surface density profiles are color-coded with the total \hicap\ masses of the galaxies.}
    \label{fig_rotation_curves}
\end{figure*}

\begin{figure*}[t]
    \centering
    \includegraphics[width = \textwidth]{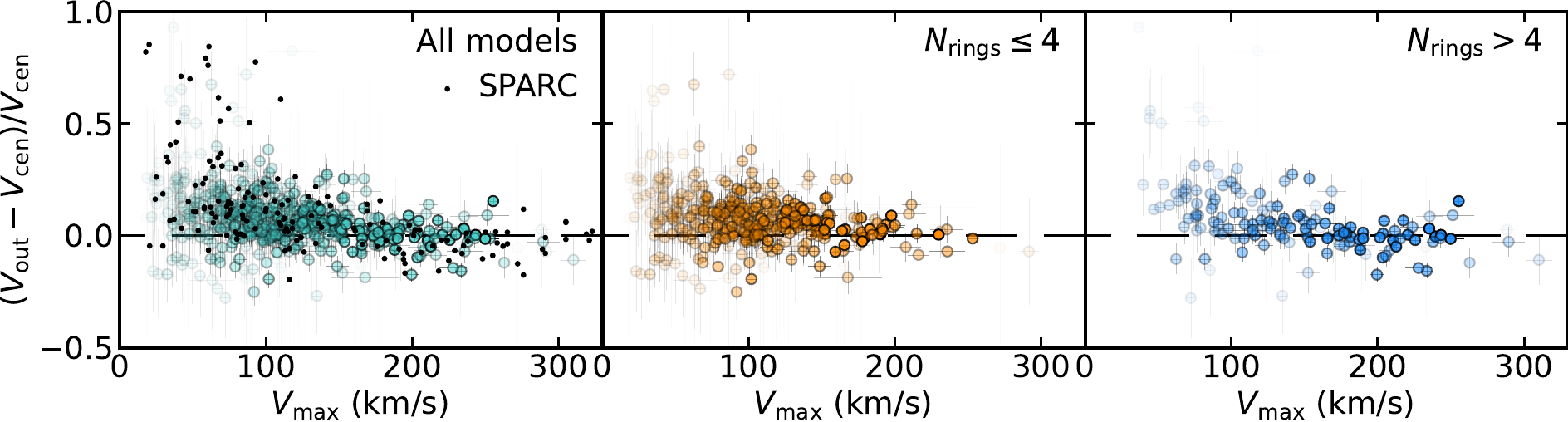}
    \caption{Outer rotation curve slope $(V_\mathrm{out} - V_\mathrm{half})/V_\mathrm{half}$ versus the maximum circular velocity $V_\mathrm{max}$ for the accepted AWES kinematic models. The transparency of the data points is proportional to the uncertainty of the slopes. In each panel, the dashed lines at a slope of zero separate rising from declining rotation curves. \textit{Left panel:} All AWES rotation curves (turquoise) compared to the SPARC rotation curves (small black dots). \textit{Middle panel:} AWES rotation curves with four or less rings. \textit{Right panel:} AWES rotation curves with more than four rings.}
    \label{fig_flatness}
\end{figure*}

Table~\ref{table_samples} also lists the rejection reasons of $A = 3$ models. The most common reason for the rejection of a kinematic model was an underestimated inclination compared to the optical images and the \hi channel maps. Of the 1772 models that were fit with the standard kinematic pipeline, 132 were assigned $A = 3$ because they should have been fit with the edge-on pipeline. These galaxies were not correctly identified as edge-on disks because of missing PGC counterparts or incorrect PGC inclinations. Similarly, 115 additional models were assigned $A = 3$ because of an otherwise severely underestimated inclination. Together, these amount to ${\sim}27\%$ of all $A = 3$ models. On the other hand, only 29 models received $A = 3$ because of an inclination estimate that is too high. For the $A = 2$ models, almost half of them missed the $A = 1$ designation because of a somewhat underestimated inclination. 

Figure~\ref{fig_acceptance} shows the distributions of PGC inclinations for models with the four acceptance flags. The mean inclination is ${\sim}\SI{50}{\deg}$ for $A = 1$ models, ${\sim}\SI{60}{\deg}$ for $A = 2$ models, and somewhat larger than that for $A = 3$ models. This trend reflects the fact that for \bbarolo{}, underestimated inclinations mainly affect highly inclined galaxies. In agreement with this, \citet{2021MNRAS.508.1195Ponomareva} find that  \bbarolo\ infers systematically lower inclinations than ellipse fits to the \SI{1}{\solarmass\per\square\parsec} contours of the moment zero maps when the latter have inclinations above ${\sim}\SI{50}{\deg}$. Furthermore, in agreement with \bbarolo\ modeling of mock \hi data \citep{2025ApJ...994..232Deg}, they find that \bbarolo\ tends to overestimate inclinations below ${\sim}\SI{40}{\deg}$. However, inclination uncertainties are already very large for such face-on galaxies because of deviations from perfectly circular disk shapes. In general, such galaxies should not be used for analyses that depend on knowledge of the overall rotation curve amplitude (as opposed to just its shape).

Observationally, an important underlying issue is the resolution and the S/N of the detections. Their effect can be seen from the large fraction of galaxies assigned $A = 3$ and 4 due to only two rings in the rotation curve (16\% and 62\%, respectively). It is also illustrated in the central and right panels of Fig.~\ref{fig_acceptance}, which show the distributions in the number of resolution elements, $N_\mathrm{res}$, and the average S/N within the \bbarolo\ masks for the four acceptance flags. The resolution, $N_\mathrm{res}$, is the spaxel area of the \bbarolo\ masks divided by the equivalent beam size at full sensitivity, $\Omega_\mathrm{beam} = \pi b_\mathrm{maj}b_\mathrm{min}/(4\ln{2})$. Kinematic modeling becomes more reliable for larger values of $N_\mathrm{res}$. We assigned $A = 1$ or 2 to no model with $N_\mathrm{res} < 3.5$ and assigned $A = 4$ to all models with $N_\mathrm{res} < 2.3$. A similar trend is present for $\langle\mathrm{S/N}\rangle$, though there is no clear difference between $A = 1$ and 2 in this case.

\subsection{Surface density profiles and rotation curves}

We give an overview of the rotation curves and surface density profiles of all 568 accepted models in Fig.~\ref{fig_rotation_curves}. They are color-coded by the total \hi mass and split into four panels, depending on the number of rings in the kinematic models. Most models, namely 234, have four rings, followed by 176 models with three rings. At higher resolutions, 59 of the models have seven or more rings with a maximum of 23 rings. Given that we adopted a common angular resolution of \SI{30}{\arcsec} for kinematic modeling, the subsamples in the panels Fig.~\ref{fig_rotation_curves} correspond to galaxies with different angular extents. These range from ${\sim}\SI{90}{\arcsec}$ in the bottom panel to more than \SI{10}{\arcmin} in the top panel. The spectral range of the AWES Cube2 data corresponds to distances from about seven to ${\sim}\SI{140}{\mega\parsec}$, so poorly resolved rotation curves in the bottom panel have extents of $2-\SI{30}{\kilo\parsec}$. The well-resolved rotation curves in the top panel reach larger maximum extents ($>\SI{60}{\kilo\parsec}$) as all panels cover roughly the same distance range. Furthermore, the extent of the \hi kinematic models correlates with $M_\mathhi$ , which is in line with the \hi mass-size relation (see Sect.~\ref{sec_hi_size_mass}).

The maximum \hi surface densities reached by the profiles lie in the range $\Sigma_\mathhi \simeq 1 - \SI{15}{\solarmass\per\square\parsec}$. Some profiles, especially at higher resolutions, show central depressions. A possible explanation is that these \hi disks are hosted by S0-Sa galaxies with large bulges \citep[e.g.,][]{2005A&A...442..137Noordermeer}. A few profiles are strong outliers toward low $\Sigma_\mathhi$ compared to the overall sample. All of these outliers have inclinations above \SI{80}{\deg}. For such high inclinations, \bbarolo{}'s method of measuring surface densities is unreliable because the surface density is measured within the highly inclined rings of the kinematic model, which are infinitely thin. Therefore, the rings cover only narrow central strips along the edge-on galaxies, which have an intrinsic thickness and are additionally broadened by the telescope's beam.

Turning to the rotation curves, the median $V_\mathrm{rot}$ in the three outer rings ranges from ${\sim}20$ to \SI{300}{\kilo\meter\per\second}. There is also an apparent correlation, albeit with significant scatter, between the amplitude of the rotation curves and $M_\mathhi$. This can be understood as a combined effect of the stellar-mass Tully-Fisher relation \citep{1977A&A....54..661Tully} and the stellar mass-\hi mass relation \citep{2012ApJ...756..113Huang, 2015MNRAS.447.1610Maddox}. Qualitatively, this agrees with the significant scatter of the $M_\mathhi$-$V_\mathrm{rot}$ relation \citep[e.g., \SI{0.74}{\dex} from][]{2024ApJ...976..159Deg}. However, a part of the scatter is certainly caused by the inclusion of low-inclination galaxies with substantial uncertainties on $V_\mathrm{rot}$ and the inclusion of galaxies with multiple distance solutions from the CF4 flow model (see Sect.~\ref{sec_dist}).

To give a more comprehensive overview of the rotation curve shapes, Fig.~\ref{fig_flatness} shows a proxy of the outer rotation curve slope against the maximum circular velocity $V_\mathrm{max}$. This proxy is $(V_\mathrm{out} - V_\mathrm{half})/V_\mathrm{half}$, where $V_\mathrm{out}$ is the circular velocity of the outermost ring and $V_\mathrm{half}$ is the circular velocity at half the radius of the outermost ring. Specifically, we derive $V_\mathrm{half}$ by linearly interpolating between the two rotation curve points that bracket this radius. We chose the central velocity instead of, for example, the first or the penultimate ring, to minimize biases from the large variety in physical resolutions of our rotation curves.

The left panel of Fig.~\ref{fig_flatness} uses all 568 AWES rotation curves (turquoise circles), where data points with larger errors in the slope are more transparent. The bulk of the AWES rotation curves has a slope close to zero, indicating that they are approximately flat in their outer halves. For low values of $V_\mathrm{max}$, there is a trend toward larger values of the outer slope ($\lesssim 0.25$). Taking $V_\mathrm{max}$ as a proxy for dynamical mass, this indicates that the rotation curves of low-mass galaxies are rising in their outer parts.

The small black points show the same quantities for the Surface Photometry and Accurate Rotation Curves \citep[SPARC;][]{2016AJ....152..157Lelli} database, which collects high-quality \hi rotation curves of 175 local galaxies. Our general trend is in good agreement with the SPARC galaxies, although we have a larger amount of seemingly declining rotation curves (slope between 0 and $-0.25$) for $V_\mathrm{max} \lesssim \SI{200}{\kilo\meter\per\second}$. However, these rotation curves do typically have large associated uncertainties and are probably driven by low S/N in the outer parts of the \hi disks. Furthermore, SPARC contains 14 galaxies with strongly rising rotation curves ($>0.5$) and low $V_\mathrm{max}$. These seem to be less represented in our sample, but the uncertainties of many of our galaxies at such values of $V_\mathrm{max}$ do allow for such slopes, hinting at the difficulty of measuring accurate rotation curve slopes in the marginally resolved regime.

The central and right panels of Fig.~\ref{fig_flatness} show the subsamples of our rotation curves with a maximum of four and more than four rings, respectively. The overall distributions are largely similar, which indicates that the general rotation curve slopes are not too strongly biased by resolution effects. Still, the trend of larger slopes for lower values of $V_\mathrm{max}$ is more recognizable for rotation curves that are better resolved. Less resolved rotation curves seem to have a significantly larger scatter in their slopes.

\begin{figure*}[t]
    \centering
    \includegraphics[width = \textwidth]{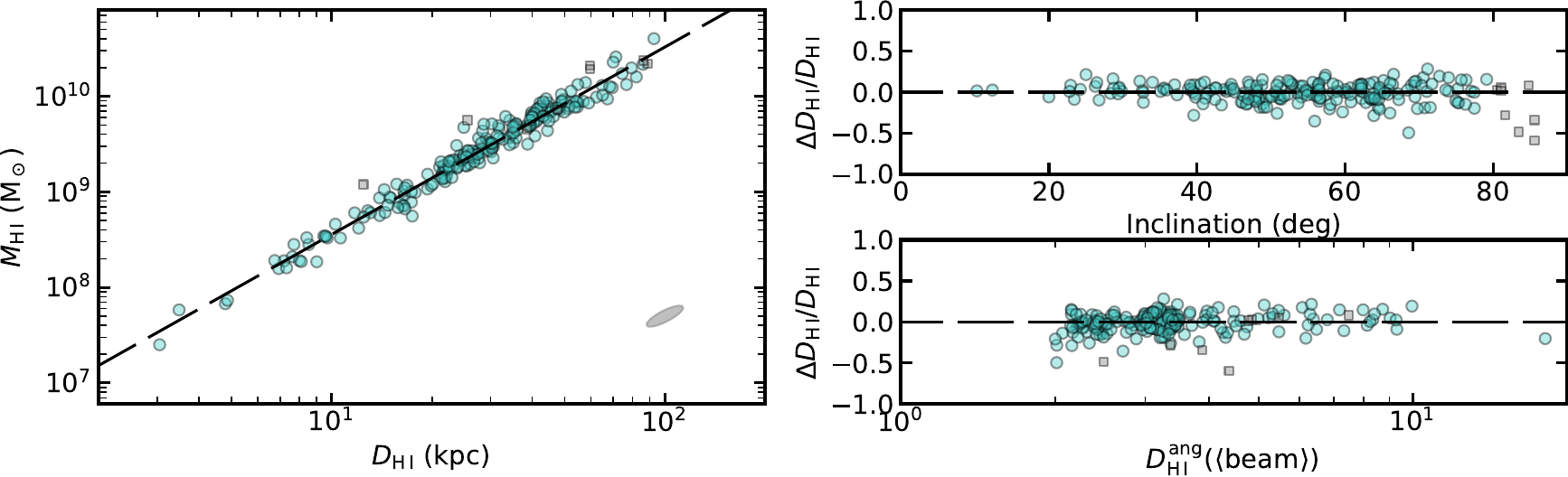}
    \caption{\hi mass-size relation of 213 galaxies from the AWES sample. The dashed black line is our best-fit relation. The gray inset in the left panel shows the median $1\sigma$ uncertainty of the data points due to flux and distance errors. \textit{Right panel:} Residuals around the best-fit relation plotted against the kinematic galaxy inclination and the \hi diameter in units of the mean beam size, $\sqrt{b_\mathrm{maj}b_\mathrm{min}}$. In all panels, the eight galaxies with inclinations above \SI{80}{\deg} are shown as gray squares. They were not included in the fit.}
    \label{fig_hims}
\end{figure*}

\begin{figure}[t]
    \centering
    \includegraphics[width = \columnwidth]{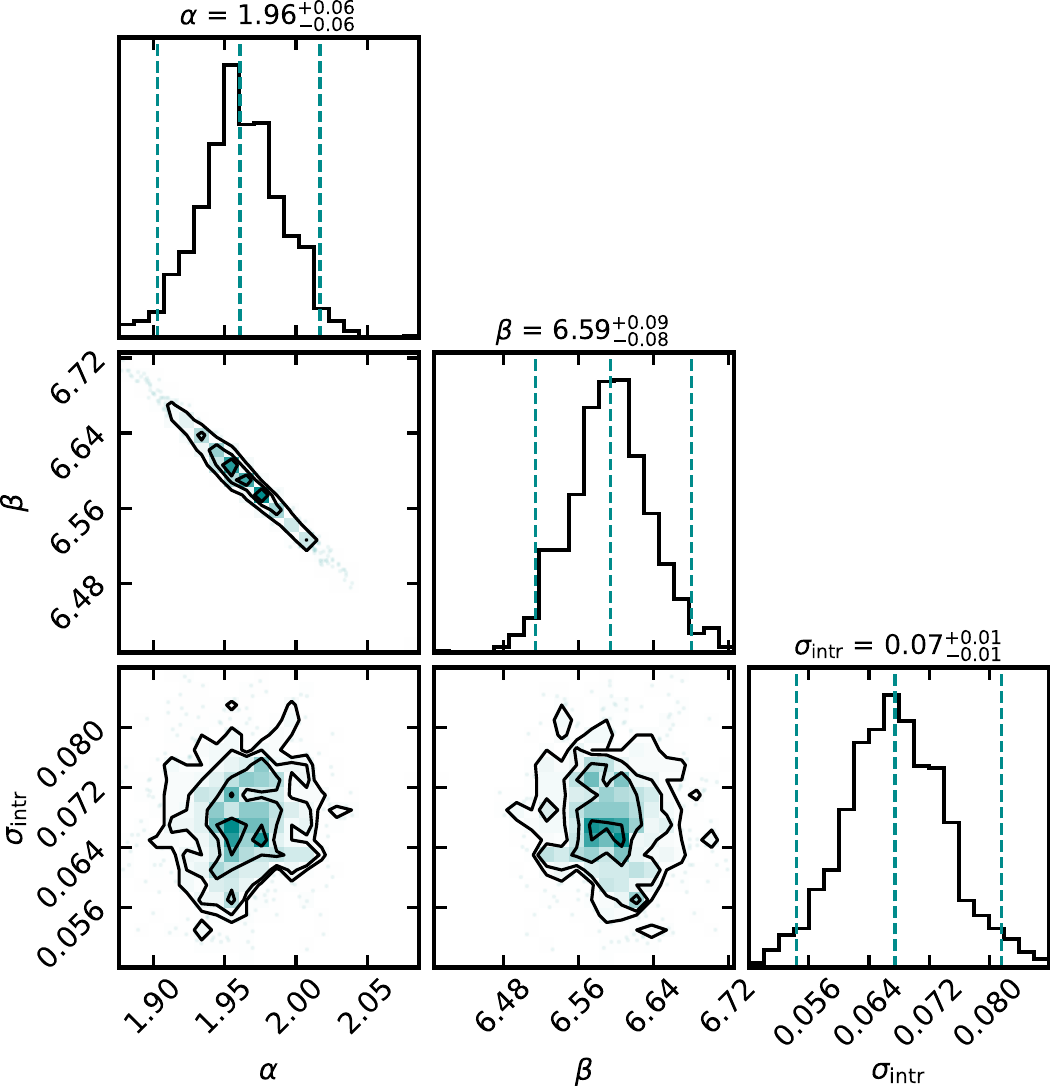}
    \caption{Posterior probability distributions of the parameters (slope, $\alpha$, intercept, $\beta$, and intrinsic scatter, $\sigma_\mathrm{intr}$) of the power law fit to the \hi\ mass-size relation. The vertical lines in the 1D histograms mark the 2.5, 50, and 97.5\% quantiles (median and $\pm2\sigma$ for a 1D Gaussian) and the contours in the 2D histograms mark the 11.8, 39.3, 67.5, and 86.4\% quantiles (0.5, 1, 1.5, and $2\sigma$ for a 2D Gaussian).}
    \label{fig_corners}
\end{figure}

\subsection{H\,{\small I} mass-size relation}  \label{sec_hi_size_mass}

The left panel of Fig.~\ref{fig_hims} shows the \hi mass-size relation obtained from the AWES sample, where $D_\mathhi$ and $M_\mathhi$ were determined as described in Sect.~\ref{sec_dist}. It contains data from 213 galaxies with reliable kinematic models ($A = 1$ and $A = 2$), for which the sensitivity is deep enough to determine $D_\mathhi$ and the resolution is good enough to correct for beam smearing. This is the largest completely homogeneous sample used until now to study the \hi mass-size relation. The correlation obtained from the AWES galaxies is very tight and spans three orders of magnitude in \hi mass.

We performed a power law fit in log-log space using a Markov chain Monte Carlo technique. Specifically, we used the Regression and Optimisation with X and Y errors code \citep[ROXY;][]{2023OJAp....6E..42Bartlett} with the marginalized normal regression method, 500 warmup steps, and 1000 samples. We infer the intrinsic scatter of the relation, where we consider the full covariance matrix with the 15\% flux uncertainty in $M_\mathhi$ and the distance uncertainties in $D_\mathhi$ and $M_\mathhi$. This specific covariance, combined with the slope of the relation close to two, is the main reason for the observed tightness, as distance errors only move galaxies along the relation (see the inset in Fig.~\ref{fig_hims}, which shows the median $1\sigma$ uncertainty range). We exclude galaxies with inclinations above \SI{80}{\deg} from the fit because they have unreliable surface density profiles and therefore underestimated $D_\mathhi$ values. They are responsible for the most prominent outliers (see the top right panel of Fig.~\ref{fig_hims}). This removes eight galaxies from the sample. The best-fit power law is
\begin{equation}    \label{eq_hims}
    \log{(M_\mathhi/\SI{}{\solarmass})} = (1.96\pm0.03)\log{(D_\mathhi/\SI{}{\kilo\parsec})} + (6.60 \pm 0.05),
\end{equation}
where the median values with the 16th and 84th percentiles are given in parentheses.

The relation has an observed vertical scatter of only \SI{0.101}{\dex} and an inferred intrinsic vertical scatter of \SI{0.067 \pm 0.007}{\dex}. This intrinsic scatter is consistent with the quadratic difference between the observed scatter and the 15\% flux uncertainty. Our best-fit slope is consistent with two, which implies a constant average surface density of the \hi disks of galaxies across three orders of magnitude in $M_\mathhi$. Our best-fit intercept gives a characteristic average surface density of ${\sim}\SI{5}{\solarmass\per\square\parsec}$ within $D_\mathhi$. However, this value is contingent on the choice of a surface density of \SI{1}{\solarmass\per\square\parsec} for defining $D_\mathhi$.

Figure~\ref{fig_corners} shows the posterior probabilities of the MCMC fit, where the typical degeneracy between the slope and intercept of a linear fit is clearly visible.  The lower right panel of Fig.~\ref{fig_hims} shows the residuals around the best-fit relation plotted against the \hi diameter $D_\mathhi^\mathrm{ang}$ in units of the mean beam size $\sqrt{b_\mathrm{maj}b_\mathrm{min}}$. A remaining bias from beam smearing would manifest itself as a trend toward larger physical \hi diameters for smaller angular diameters $D_\mathhi^\mathrm{ang}$. No such trend is visible. We also performed the fit without any beam smearing corrections, in which case we find a slope of $1.99 \pm 0.03$, an intercept of $6.51\pm0.04$, and an intrinsic scatter of $0.072\pm0.007$. The main effect of the beam smearing correction is to increase the intercept of the relation, as it reduces the \hi diameters.

Most studies in the last decade write the \hi mass-size relation as $D_\mathhi \propto {M_\mathhi}^\alpha$, so we also perform our fit with that axis convention. In this way, we find
\begin{equation}
    \log{(D_\mathhi/\SI{}{\kilo\parsec})} = (0.501 _{-0.008}^{+0.007})\log{(M_\mathhi/\SI{}{\solarmass})} - (3.28_{-0.06}^{+0.08}),
\end{equation}
with an observed vertical scatter of \SI{0.047}{\dex} and an intrinsic vertical scatter of \SI{0.034\pm0.004}{\dex}. These values agree with our fit for the $M_\mathhi \propto {D_\mathhi}^\alpha$ convention when we convert them back into this shape.

Turning to previous studies, our best fit is in perfect agreement with that of \citet{2022MNRAS.512.2697Rajohnson}, who used the largest homogeneous sample previous to our own (204 galaxies from MIGHTEE-HI). They find  a slope of $0.501 \pm 0.008$ and an intercept of $-3.252_{-0.074}^{+0.073}$. Furthermore, our best fit is consistent with that of \citet{2016MNRAS.460.2143Wang}. These authors used the largest sample so far, namely 501 unique galaxies compiled from different surveys, and found a slope and intercept of $0.506 \pm 0.003$ and $-3.293 \pm 0.009$, respectively. However, \citet{2024ApJ...976..159Deg} find a larger slope of $0.533 \pm 0.007$ and a lower intercept of $-3.55 \pm 0.07$ with 148 galaxies from WALLABY. This apparent discrepancy is likely driven by the degeneracy between slope and intercept of a linear fit (see Fig.~\ref{fig_corners}). In fact, all previous studies that found a slope larger than $0.52$ also found an intercept smaller than $-3.5$ \citep{2001A&A...370..765Verheijen, 2002A&A...390..829Swaters, 2005A&A...442..137Noordermeer, 2016A&A...585A..99Martinsson, 2016AJ....152..157Lelli}.

With an observed scatter of ${\sim}\SI{0.05}{\dex}$ in the $D_\mathhi$ direction, the tightness of our relation lies between that of \citet[\SI{0.04}{\dex}]{2024ApJ...976..159Deg} and those of \citet[\SI{0.06}{\dex}]{2016MNRAS.460.2143Wang} and \citet[\SI{0.057}{\dex}]{2022MNRAS.512.2697Rajohnson}. Our main new result is an inference of the intrinsic vertical scatter of the \hi mass-size relation of \SI{0.067\pm0.007}{\dex} in the $M_\mathhi$ direction and \SI{0.034\pm0.004}{\dex} in the $D_\mathhi$ direction. This intrinsic scatter was also studied by \citet{2022MNRAS.512.2697Rajohnson}, who find a larger value of \SI{0.054\pm0.003}{\dex} in the $D_\mathhi$ direction, comparable to their observed scatter of \SI{0.057}{\dex}. They interpret this as possible evidence for an intrinsic ${\sim}10\%$ variation of $D_\mathhi$ at fixed $M_\mathhi$. The most probable reason for the discrepancy between their result and our own are differences in the observational uncertainty estimation. Although \citet{2022MNRAS.512.2697Rajohnson} propagate inclination and beam size uncertainties into the uncertainty of $D_\mathhi$, which we do not do, the largest difference comes from the fact that they estimate the flux uncertainty from the cube noise in line-free regions. They cite typical \hi mass uncertainties
of ${\sim}5\%$ above $M_\mathhi \simeq \SI{e9}{\solarmass}$ and of ${\sim}10\%$ between ${\sim}\SI{e8}{}$ and ${\sim}\SI{e9}{\solarmass}$. They do not consider distance uncertainties on $M_\mathhi$, which do not significantly enter the \hi mass-size relation anyway.

Our \hi flux uncertainties of 15\% across all \hi masses are notably larger. They include errors from the Apertif flux calibration, mosaicking, and primary beam correction \citep{2022A&A...667A..39Kutkin} and are in line with the uncertainty estimated by comparing to single-dish fluxes (see Appendix~\ref{app_flux}). As a rough test of this explanation of the discrepancy, we repeat our MCMC fit with a fixed constant uncertainty on $M_\mathhi$ of 7\% and without covariances \citep[as in the fit of][]{2022MNRAS.512.2697Rajohnson}. With this setup, we infer an intrinsic scatter of \SI{0.045 \pm 0.003}{\dex}, close to the MIGHTEE-HI result and consistent with our observational scatter. Nevertheless, full agreement cannot be achieved as our observed scatter is ${\sim}\SI{0.01}{\dex}$ lower than theirs. It seems unlikely that this difference stems from a genuine difference in the intrinsic scatter of the samples. While \citet{2022MNRAS.512.2697Rajohnson} trace a larger range in redshift ($z < 0.084$), they do not report significant differences in scatter between different redshift bins. Instead, the observational errors in our sample might be lower, possibly because of our slightly better spatial resolutions (a median of four beams along the major axis compared to a median of three beams).

Finally, the \hi mass-size relation acts as a powerful diagnostic for simulations of galaxy formation. Modern simulations are able to reproduce the general shape of the relation quite well \citep[e.g.,][]{2018MNRAS.476.3744Lutz, 2019MNRAS.487.1529Diemer}, unless their feedback prescriptions are so strong that they disrupt the \hi disks \citep{2016MNRAS.456.1115Bahe}. In the future, an additional challenge for theoretical models will be to reproduce the small intrinsic scatter of the relation. As an example, an investigation with the AURIGA simulation shows that the overall shape of the relation can be reproduced, but its intrinsic scatter is ${\sim}\SI{0.12}{\dex}$ \citep{2017MNRAS.466.3859Marinacci}, much larger than inferred both by us and by \citet{2022MNRAS.512.2697Rajohnson}.

\section{Summary}   \label{sec_summary}

We present the first release of rotation curves and surface density profiles from the AWES survey conducted with the Apertif system on the WSRT. These kinematic models are based on 1924 \hi data cubelets (after an initial S/N cut of 3.5 to the cubelet voxels) and were automatically derived with the \bbarolo\ kinematic modeling code. We assessed the reliability of all kinematic models through visual inspection. The main results are as follows:
\begin{itemize}
    \item Automated modeling with \bbarolo\ resulted in reliable kinematic models for 568 of the 1924 galaxies (${\sim}30\%$). These models form the basis of our data release. Rejected models mostly suffer from incorrect estimations of the geometric parameters of their disks; in particular, underestimated inclinations. They tend to have low average S/N values and low resolution, driving the uncertainty of geometric parameter estimates. 
    \item The galaxies with reliable models span a large range in parameter space. They have extents between ${\sim}2$ and more than \SI{60}{\kilo\parsec}, rotation velocities between ${\sim}20$ and ${\sim}\SI{300}{\kilo\meter\per\second}$, and \hi masses between \SI{2.5e7}{} and \SI{4e10}{\solarmass}.
    \item Most kinematic models have four rings (234 galaxies) or three rings (176 galaxies). There are 158 models with at least five rings, with a tail up to 23 rings.
    \item Most rotation curves become flat at large radii; in particular, those of massive galaxies with large maximum circular velocities. However, many rotation curves of low-mass galaxies keep rising in their outer parts, although they also show more apparent scatter in their slopes. This scatter is at least partially driven by the typically lower physical resolutions of these galaxies.
\end{itemize}

Furthermore, we used the surface density profiles measured by \bbarolo\ to study the \hi mass-size relation of 213 AWES galaxies for which the \hi diameter, $D_\mathhi$, could be measured at \SI{1}{\solarmass\per\square\parsec}. This is the largest homogeneous sample ever used to study this relation. Our findings are as follows:
\begin{itemize}
    \item Writing the relation as $M_\mathhi \propto {D_\mathhi}^\alpha$, we find a slope of $1.96\pm0.03$ and an intercept of \SI{6.60\pm0.05}{} of the relation (16th, 50th, and 84th percentiles), in agreement with previous measurements \citep{2016MNRAS.460.2143Wang, 2022MNRAS.512.2697Rajohnson}.
    \item The relation is very tight with an observed scatter in the $M_\mathhi$ direction of \SI{0.101}{\dex}. We infer an intrinsic scatter in the same direction of \SI{0.067\pm 0.007}{\dex}.
    \item Our intrinsic scatter is smaller than the value determined by \citet[\SI{0.108\pm 0.006}{\dex} when converted to the $M_\mathhi$ direction]{2022MNRAS.512.2697Rajohnson}. The disagreement can largely be explained by our larger \hi flux uncertainty estimates (see Appendix~\ref{app_flux}), but a remaining discrepancy seems to be related to our somewhat lower observed scatter. While the intrinsic scatter of scaling relations contains critical additional information for testing theoretical models beyond their shape, our result underlines that such inferences hinge critically on accurate observational uncertainty estimates.
\end{itemize}

The rotation curves and surface density profiles provided here constitute the largest homogeneous database of its type up until now. Although the mean spatial resolution is lower than for other state-of-the-art databases such as SPARC \citep{2016AJ....152..157Lelli}, the data are more homogeneous and the sample size will allow for novel statistical, population-level investigations (e.g., into environmental dependencies or the resolved galaxy velocity function). In the future, 400 galaxies drawn from the Apertif sample will be complemented with integral field spectroscopy data from the WEAVE-Apertif project \citep{2025hsa..conf...82FalconBarroso}, which will allow for the derivation of higher resolution rotation curves in the inner parts of these galaxies. In terms of \hi science, our visually inspected kinematic models provide a testbed for developing automated kinematic modeling pipelines, which will be crucial for exploiting the vast amounts of \hi data from future large-scale surveys with the SKA Project and the DSA \citep{2019BAAS...51g.255Hallinan}.

\section*{Data availability}

The kinematic models and their overview plots are available at the CDS via \href{https://cdsarc.cds.unistra.fr/viz-bin/cat/J/A+A/714/A250}{https://cdsarc.cds.unistra.fr/viz-bin/cat/J/A+A/714/A250}.

\begin{acknowledgements}
    KH thanks Thijs van der Hulst, Tom Oosterloo, and in particular the anonymous referee for helpful comments that improved the quality of this manuscript. EDT acknowledges support by the European Research Council (ERC) under grant agreement no.\ 101040751. WJGdB acknowledges funding from the ERC under the European Union’s Horizon 2020 research and innovation programme (grant agreement No 882793 ‘MeerGas’). JG, MP, and SSE acknowledge financial support from the grants PID2021-123930OB-C21 and PID2024-155817OB-I00 funded by the Ministerio de Ciencia, Innovación y Universidades (MICIU) / Agencia Estatal de Investigación (AEI) / 10.13039/501100011033, and by the European Regional Development Fund (ERDF) / European Union, the grant CEX2021-001131-S funded by MICIU/AEI/10.13039/501100011033, and the grant INFRA24023 (CSIC4SKA) funded by Consejo Superior de Investigaciones Científicas (CSIC). KMH, JG, MP, and SSE acknowledge the Spanish Prototype of an SRC \citep[SPSRC;][]{2022JATIS...8a1004Garrido} service and support funded by MICIU, by the Junta de Andalucía, by the ERDF and by the European Union NextGenerationEU/PRTR. The SPSRC acknowledges financial support from AEI through the "Center of Excellence Severo Ochoa" award to the Instituto de Astrofísica de Andalucía (IAA-CSIC) (SEV-2017-0709) and from the grant CEX2021-001131-S funded by MICIU/AEI/ 10.13039/501100011033. KMH and KH acknowledge support from the Sweden SKA Regional Center (sweSRC) node operated by Onsala Space Observatory in collaboration with Chalmers e-Commons. The Onsala Space Observatory national research infrastructure is funded through Swedish Research Council grant No 2019-00208. JvL was supported by CORTEX (NWA.1160.18.316), under the research programme NWA-ORC, financed by the Dutch Research Council ("De Nederlandse Organisatie voor Wetenschappelijk Onderzoek", NWO). This works makes use of data from the Apertif system installed at the Westerbork Synthesis Radio Telescope owned by ASTRON. ASTRON, the Netherlands Institute for Radio Astronomy, is an institute of the NWO. This research has made use of the NASA/IPAC Extragalactic Database (NED) which is operated by the Jet Propulsion Laboratory, California Institute of Technology, under contract with the National Aeronautics and Space Administration. This work used images from the DESI Legacy Imaging Surveys, the Sloan Digital Sky Survey, and the Pan-STARRS1 survey. Their full acknowledgments can be found at \href{https://www.legacysurvey.org/acknowledgment/}{https://\-www.legacysurvey.org/\-acknowledgment/}, \href{https://www.sdss.org/collaboration/citing-sdss/}{https://\-www.sdss.org/\-collaboration/\-citing\--sdss/}, and \href{https://outerspace.stsci.edu/spaces/PANSTARRS/overview/}{https://\-outerspace.stsci.edu/\-spaces/\-PANSTARRS/\-overview/}, respectively.
\end{acknowledgements}

\bibliographystyle{aa}
\bibliography{references}

\begin{appendix}

\section{Continuum oversubtraction} \label{app_baseline}

\begin{figure*}
    \centering
    \includegraphics[width = \textwidth-.75cm]{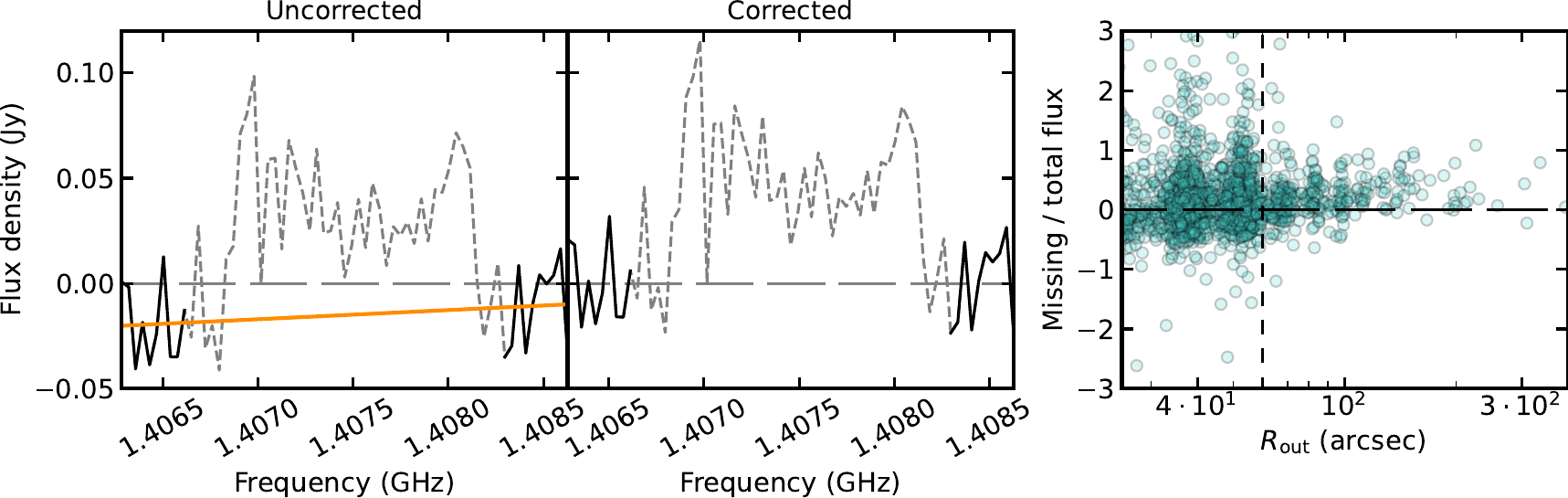}
    \caption{Illustration of our correction for continuum oversubtraction in the AWES cubelets. \textit{Left panel:} Full uncorrected cubelet spectrum for NGC\,1171. The dashed gray line indicates a flux of zero. The solid black lines are the spetrum in the line-free channels used for fitting a linear function (orange) to the continuum. \textit{Middle panel:} Full spectrum of the corrected NGC\,1171 cubelet (dashed gray line as before). \textit{Right panel:} The ratio of missing flux due to oversubtraction to total uncorrected cubelet flux plotted against the radius of the outermost kinematic model ring. The x-axis has an additional jitter of \SI{3}{\arcsec}. We corrected for oversubtraction in galaxies to the right of the vertical dashed line (\SI{60}{\arcsec}).}
    \label{fig_baselines}
\end{figure*}

A subset of the AWES cubelets is affected by oversubtracted continuum emission, which results in negative spectral baselines. The full spectrum of the cubelet of AHCJ030358.2+432358 (NGC\,1171), one of the most strongly affected galaxies, can be seen in the left panel of Fig.~\ref{fig_baselines}. In general, this has a minor effect on our kinematic modeling pipeline, but it can result in underestimated \hi fluxes. More extended sources are more strongly affected and the resulting bias can, for example, become notable in the \hi mass-size relation.

A more detailed discussion of this effect will follow in Hess et al. (in prep.). Here, we corrected our cubelets by first fitting a linear function (orange in Fig.~\ref{fig_baselines}) to the line-free channels (solid black) around our galaxies. Based on the chosen SOFIA parameters, these channels are ten at either spectral end of the cubelets. For a small number of galaxies that lie at one of the spectral edges of their cubelets, we used the unaffected opposite ten channels and replaced the linear function with a constant one to account for the reduced constraining power of the data.  We then corrected the cubelets' global spectra by adding the missing flux estimated from the linear fit on a channel-by-channel basis. The middle panel of Fig.~\ref{fig_baselines} shows the full cubelet spectrum for NGC\,1171 after this correction.

We do not apply this correction to all cubelets as the linear fit can be uncertain in some cases. The right panel of Fig.~\ref{fig_baselines} shows the missing flux estimated from integrating the linear fits along the cubelets' spetral extent divided by the total summed flux within the uncorrected cubelets. The missing flux has a positive sign for negative baselines and vice versa. As the denominator is the uncorrected flux, larger positive values correspond to missing fluxes that approach the corrected fluxes of the respective galaxies. The ratio is plotted against the outermost angular extent, $R_\mathrm{out}$, of the kinematic models from a previous run of our pipeline on the uncorrected cubelets. The scatter decreases significantly above $R_\mathrm{out}\sim\SI{60}{\arcsec}$ (vertical dashed lines), which is also where a systematic deviation to negative baselines becomes apparent. Therefore, we corrected only the 291 cubelets to the right of this line. We use a flag in our data release to indicate which cubelets were corrected.

\section{Flux accuracy} \label{app_flux}

\begin{figure*}
    \centering
    \includegraphics[width = \textwidth-.75cm]{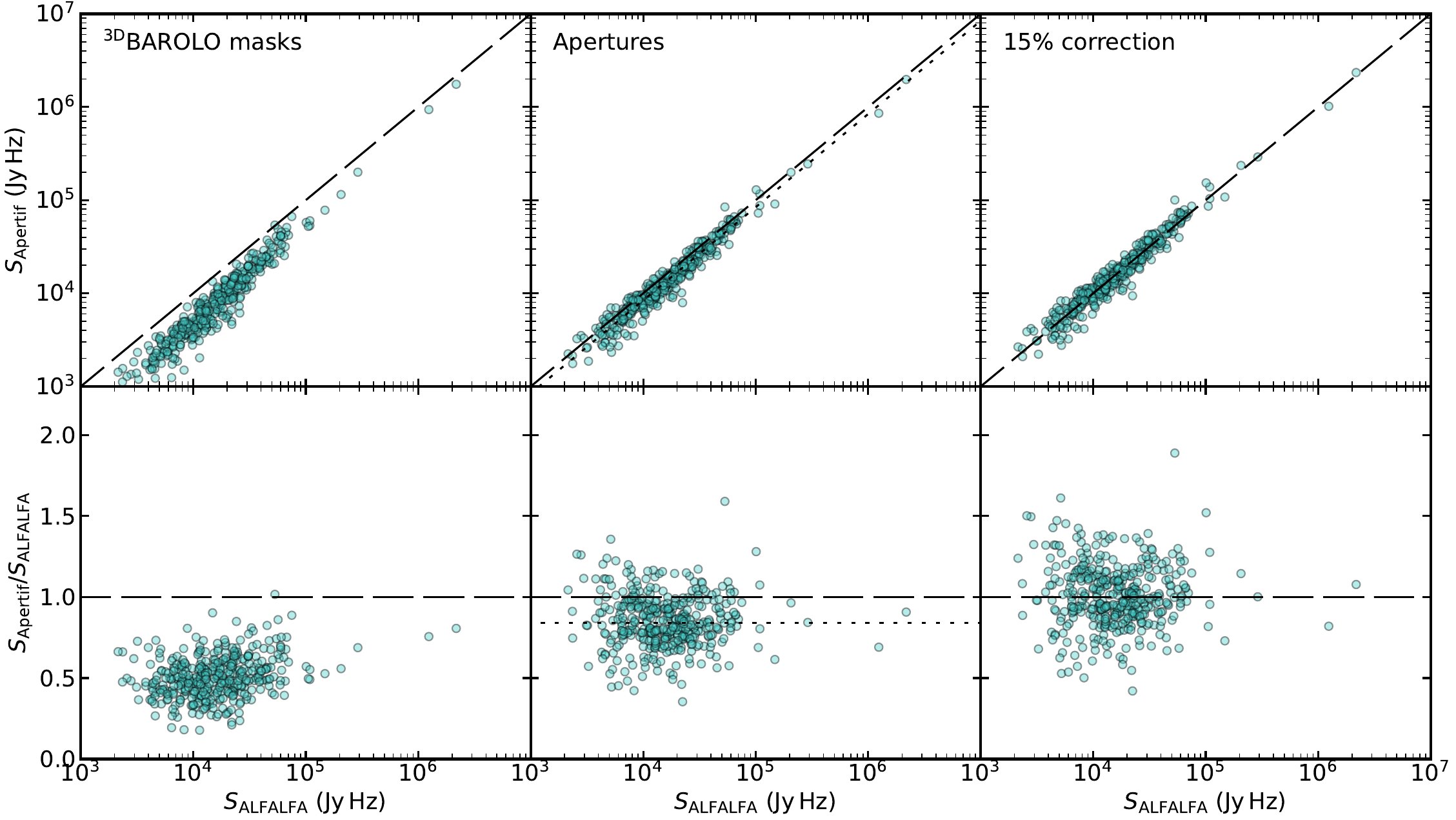}
    \caption{Comparison of Apertif fluxes to ALFALFA fluxes. The top row shows the absolute Apertif fluxes and the bottom row the ratio of Apertif to ALFALFA measurements. Long dashed lines indicate equality between the two surveys. \textit{Left panels:} The Apertif fluxes measured within the \bbarolo\ masks show a flux-dependent underestimation compared to the single-dish values. \textit{Middle panels:} Measuring Apertif fluxes within circular apertures with a radius \SI{30}{\arcsec} larger than the extent of the \bbarolo\ models removes the flux-dependent trend and significantly decreases the discrepancy with ALFALFA. However, there remains a constant offset between the Apertif fluxes (dotted line) and the ALFALFA ones. \textit{Right panels:} Multiplying the Apertif fluxes with a constant factor to correct for the missed 15\% flux restores agreement with the single-dish measurements.}
    \label{fig_fluxes}
\end{figure*}

Figure~\ref{fig_fluxes} shows the comparison of the Apertif fluxes, $S_\mathrm{Apertif}$, of 421 galaxies with fluxes $S_\mathrm{ALFALFA}$ from the single-dish ALFALFA survey \citep{2018ApJ...861...49Haynes}. We converted ALFALFA fluxes, which are given in \SI{}{\jansky\kilo\meter\per\second} using the optical velocity convention, to \SI{}{\jansky\hertz} \citep[Eq.~41]{2017PASA...34...52Meyer}. The top row shows comparisons of the absolute values and the bottom row the ratio between the Apertif and ALFALFA fluxes. The left panels give the comparison of ALFALFA with the Apertif fluxes measured inside the masks produced by the \bbarolo\ \texttt{SEARCH} task (see Sect.~\ref{sec_prep}). The fluxes within these masks are clearly biased toward lower values compared to ALFALFA. More importantly, the bias seems not to be constant, with progressively larger discrepancies for smaller flux values. A likely explanation is that \bbarolo\ is masking galactic regions with low S/N and that this preferentially affects galaxies with overall lower fluxes and S/N. Indeed, our \bbarolo\ masks are not optimized for measuring \hi fluxes, but are meant to identify regions with high enough S/N for robust kinematic modeling.

The central panels show a comparison of ALFALFA to Apertif fluxes that were derived in circular apertures centered on the galaxies. We chose the radii of these apertures to be \SI{30}{\arcsec} larger than the outermost ring radius of the \bbarolo\ kinematic models. Larger radii only make a marginal difference to the results. This procedure results in measurements that are in better agreement with ALFALFA and do no longer show a clear flux-dependent trend. However, there is still a constant offset left. Specifically, the median value of the Apertif aperture fluxes are ${\sim}15\%$ lower than the ALFALFA values. This discrepancy is probably a result of missing short baselines in WSRT, the flux calibration of Apertif, primary beam attenuation, or a combination of these effects.

Finally, the right panels show the result of multiplying the Apertif aperture fluxes by a constant factor of 1.19\footnote{The missing 15\% are calculated relative to the ALFALFA fluxes, which corresponds to a factor of 1.19 that must be multiplied with the smaller Apertif fluxes.}. We use these corrected fluxes in our analyses. After the correction, the difference between the fluxes $\log{S_\mathrm{Apertif}} - \log{S_\mathrm{ALFALFA}}$ has a median with 16th and 84th percentile ranges of $0.003_{-0.08}^{+0.07}\,\SI{}{\dex}$. For estimating the random error $\delta S_\mathrm{Apertif}$ on the Apertif fluxes, we assume that it is comparable to the random error on the ALFALFA fluxes. We can then use the ${\sim}\SI{0.08}{\dex}$ scatter of the flux difference to arrive at $\delta S_\mathrm{Apertif} \simeq \SI{0.08}{\dex}/\sqrt{2} = \SI{0.056}{\dex}$, which corresponds to a ${\sim}13\%$ uncertainty. This is compatible with the expected Apertif flux uncertainty of 15\% \citep{2022A&A...667A..39Kutkin} and we adopt this later value for consistency with other Apertif studies \citep[e.g.,][]{2025A&A...703A.295Siljeg}.

\section{Apertif spectral response} \label{app_lsf}

The observed line broadening in an \hi data cube is a combination of the intrinsic velocity dispersion of the \hi gas, beam smearing effects due to the point spread function, and the spectral response or line spread function (LSF) of the radio interferometer. \bbarolo\ assumes that the LSF follows a Gaussian profile, so the standard deviation of the instrumental broadening, $\sigma_\mathrm{inst}$, can simply be added in quadrature with the intrinsic gas velocity dispersion, $\sigma_V$. This is a computationally less expensive alternative to the otherwise necessary convolution of the intrinsic line profiles with the LSF during the construction of the model data cubes. The instrumental broadening is passed to \bbarolo\ through the \texttt{LINEAR} parameter, which is the standard deviation of the Gaussian instrumental response in channel units.

During data processing, the original Apertif spectral channels of width \SI{2.6}{\kilo\meter\per\second} were rebinned, resulting in an effective boxcar-shaped spectral response function of width \SI{7.7}{\kilo\meter\per\second}. In general, the convolution of a Gaussian intrinsic line profile with such a boxcar will result in a combination of two error functions that approximate a Gaussian for $\sigma_V \gg \sigma_\mathrm{inst}$ and a boxcar for $\sigma_V \ll \sigma_\mathrm{inst}$. If the Gaussian approximation to this convolution is justified, an approximate \texttt{LINEAR} parameter can be calculated as
\begin{equation}
    \texttt{LINEAR}^2 = \sigma_\mathrm{inst}^2 = \sigma_\mathrm{approx}^2 - \sigma_V^2,
\end{equation}
with $\sigma_\mathrm{approx}$ the standard deviation of the Gaussian function that approximates the true convolved profile. In the future, this convolution option will be added to the \bbarolo\ \texttt{3DFIT} task, but here, it is important to test the limits of the Gaussian approximation.

\begin{figure*}
    \centering
    \includegraphics[width = \textwidth-.5cm]{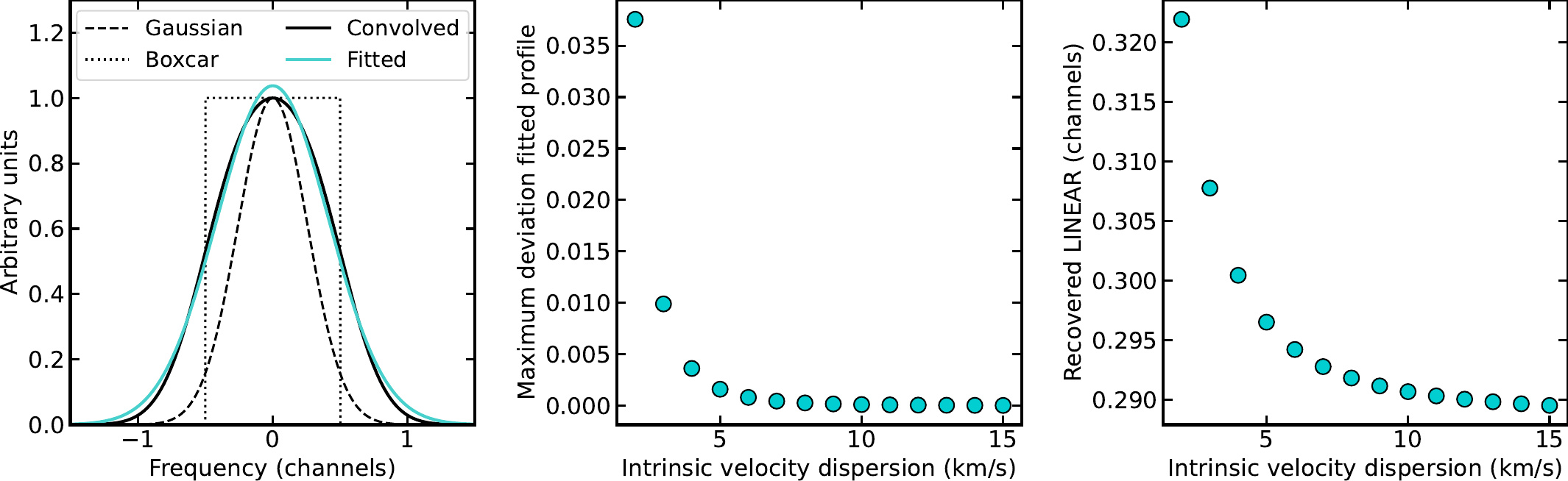}
    \caption{Illustration of the Gaussian approximation for the boxcar Apertif spectral response function. \textit{Left panel:} We convolved a Gaussian intrinsic line profile (dashed line) with a boxcar spectral response function (dotted profile). The result of the convolution is given by the solid black line and the Gaussian fit to this by the turquoise line. \textit{Central panel:} Maximum difference between the true convolved profile and the Gaussian fit for different assumed intrinsic velocity dispersions. The width of the boxcar was fixed to \SI{7.7}{\kilo\meter\per\second}, corresponding to the Apertif channel width. \textit{Right panel:} Recovered \texttt{LINEAR} parameters derived from the Gaussian approximation for different intrinsic velocity dispersions.}
    \label{fig_lsf}
\end{figure*}

For this purpose, we assume Gaussian intrinsic \hi velocity dispersions with standard deviations ranging from 2 to \SI{15}{\kilo\meter\per\second}. We convolve these with boxcar functions with a width of \SI{7.7}{\kilo\meter\per\second}, corresponding to the Apertif spectral response function. The left panel of Fig.~\ref{fig_lsf} demonstrates this procedure for $\sigma_V = \SI{2}{\kilo\meter\per\second}$, that is, where the Gaussian approximation is least reliable. Nevertheless, a Gaussian fit (turquoise) provides a good description of the true convolved profile (solid black). This is also illustrated by the central panel, which plots the maximum difference between the fit Gaussian profiles and the true convolved profiles against the intrinsic velocity dispersion. For the Apertif spectral resolution, deviations reach a level of only ${\sim}3.5\%$ compared to the true profile's peak value  when $\sigma_V = \SI{2}{\kilo\meter\per\second}$. For larger intrinsic velocity dispersions, the deviations approach zero.

Consequently, the Gaussian approximation is justified for the AWES data. The right panel of Fig.~\ref{fig_lsf} shows the recovered \texttt{LINEAR} parameters versus the intrinsic velocity dispersion. The parameter converges to a value of ${\sim}0.29$ channels, but even for $\sigma_V = \SI{2}{\kilo\meter\per\second}$, it has only a slightly larger value of 0.32. Such small differences have a minor effect on the kinematic modeling results, so we assumed $\texttt{LINEAR} = 0.3\,\mathrm{channels}$ throughout our kinematic pipeline.

\section{Kinematic model overview plots}    \label{app_plots}

\begin{figure*}[p]
    \centering
    \includegraphics[width = \textheight - 40pt, angle = 90]{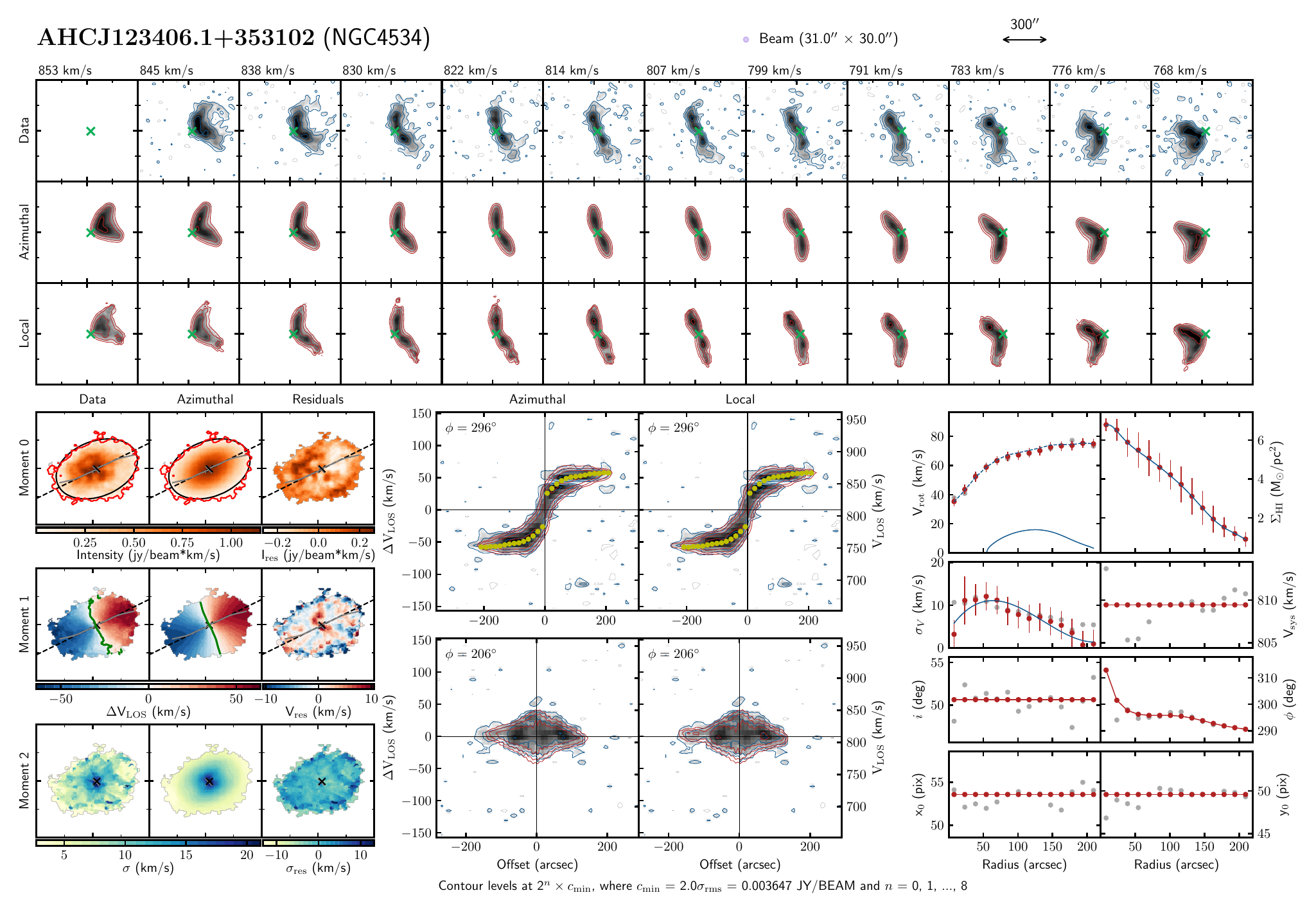}
    \caption{\bbarolo\ output for an $A = 1$ kinematic model. Panels are as in Fig.~\ref{fig_a1_plot1}. With 14 data points in the rotation curve, this is one of the more extended models in our sample. The channel maps reveal a flagged channel at the receding edge of the galaxy, which however does not noticeably affect the kinematic modeling.}
    \label{fig_a1_plot2}
\end{figure*}

\begin{figure*}[p]
    \centering
    \includegraphics[width = \textheight - 40pt, angle = 90]{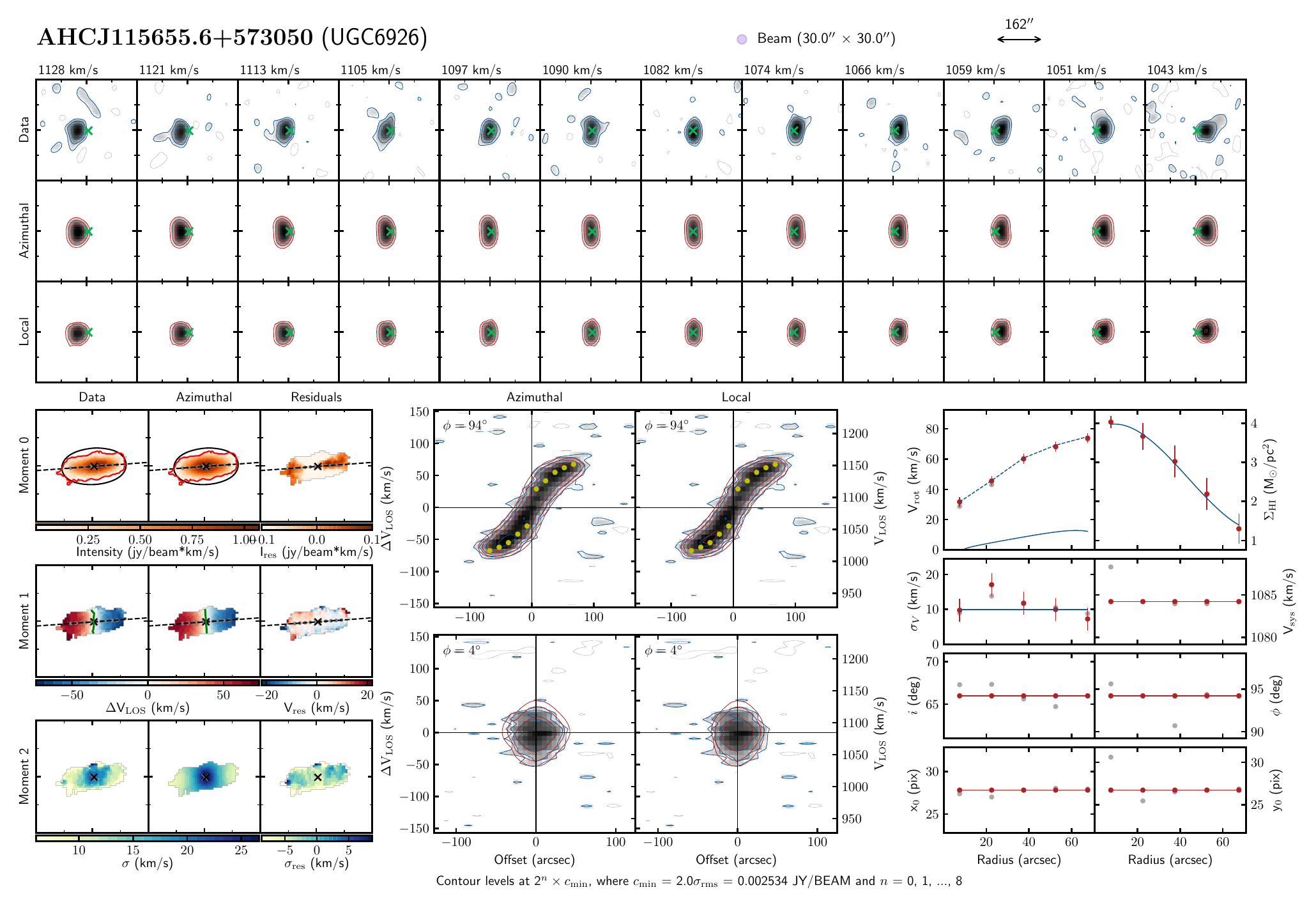}
    \caption{\bbarolo\ output for an $A = 2$ kinematic model. Panels are as in Fig.~\ref{fig_a1_plot1}. As can be seen from the moment maps and the channel maps of the azimuthally normalized model, the inclination is underestimated by ${\sim}\SI{10}{\deg}$. However, for an almost edge-on galaxy this has only a minor impact on the rotation curve.}
    \label{fig_a2_plot1}
\end{figure*}

\begin{figure*}[p]
    \centering
    \includegraphics[width = \textheight - 40pt, angle = 90]{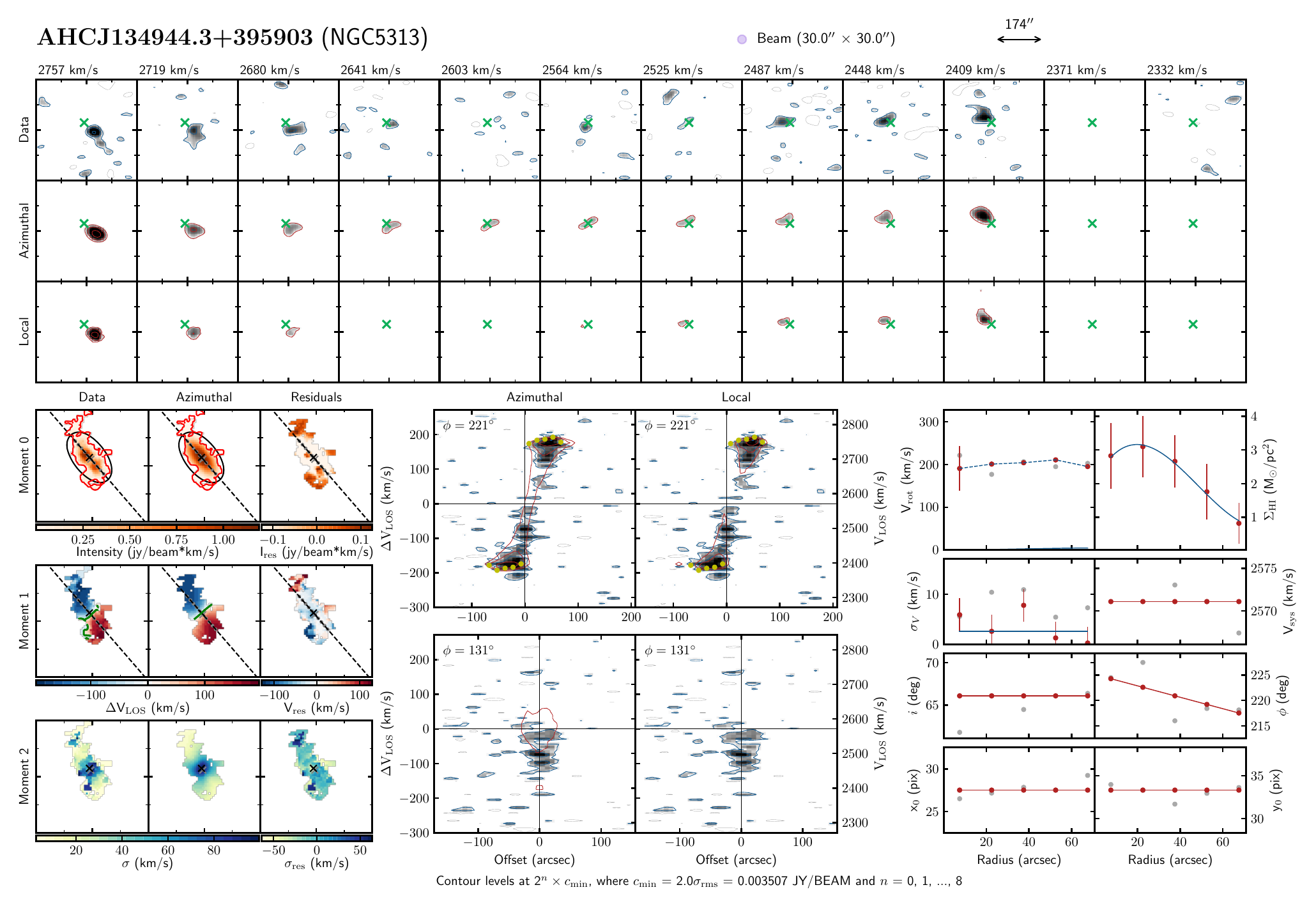}
    \caption{\bbarolo\ output for an $A = 2$ kinematic model. Panels are as in Fig.~\ref{fig_a1_plot1}. \bbarolo\ incorrectly changed the kinematic center (black in the moment maps) compared to the optical value (gray), but the impact on the shape of the rotation curve is minor. The reason for the wrong center is likely related to an imperfect estimation of the systemic velocity because of the flagged channel at \SI{2371}{\kilo\meter\per\second}.}
    \label{fig_a2_plot2}
\end{figure*}

Figure~\ref{fig_a1_plot2} shows an additional example of the \bbarolo\ overview plots for an $A = 1$ kinematic model, namely for the galaxy AHCJ123406.1+353102 (NGC\,4534). Figures~\ref{fig_a2_plot1} and \ref{fig_a2_plot2} show the overview plots of two $A = 2$ kinematic models. The first of the two, for the galaxy AHCJ115655.6+573050 (UGC~6926), has a somewhat too low inclination in the kinematic model, by roughly \SI{10}{\deg}. This can be seen from the moment zero maps, where the black ellipse and the red line trace the same surface density in the model and data, respectively, and from the channel maps of the azimuthally normalized model. However, at an inclination of ${\sim}\SI{70}{\deg}$, an inclination error of ${\sim}\SI{10}{\deg}$ changes $V_\mathrm{rot}$ by less than 10\%. In the case of the second $A = 2$ model in Figure~\ref{fig_a2_plot2} (AHCJ134944.3+395903, NGC\,5313), the kinematic center (black in the moment maps) was incorrectly moved away from the optical value (gray) and the systemic velocity is somewhat too large. This is also visible in the PVDs. The reason for this is likely related to the flagged channel that is visible on the approaching side in the major axis PVD and from the empty panel in the channel maps. Nevertheless, the impact of this error is small considering the flatness of the rotation curve, the large uncertainty in the innermost ring, and the large rotation velocity.

\begin{figure*}[p]
    \centering
    \includegraphics[width = \textheight - 40pt, angle = 90]{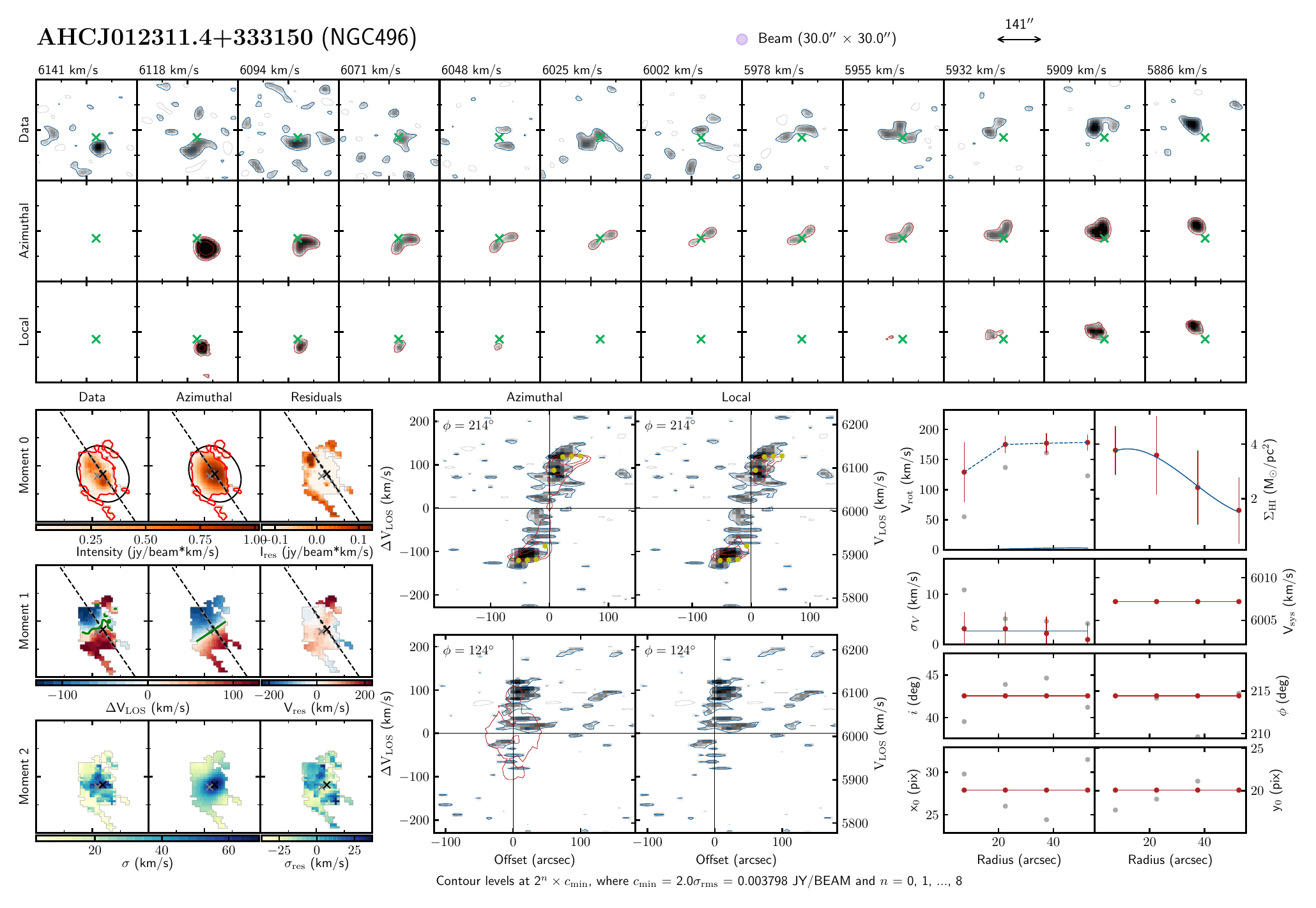}
    \caption{\bbarolo\ output for an $A = 3$ kinematic model. Panels are as in Fig.~\ref{fig_a1_plot1}. \bbarolo\ converges to an incorrect kinematic center for the galaxy, which renders the resulting kinematic model unreliable.}
    \label{fig_a3_plot1}
\end{figure*}

\begin{figure*}[p]
    \centering
    \includegraphics[width = \textheight - 40pt, angle = 90]{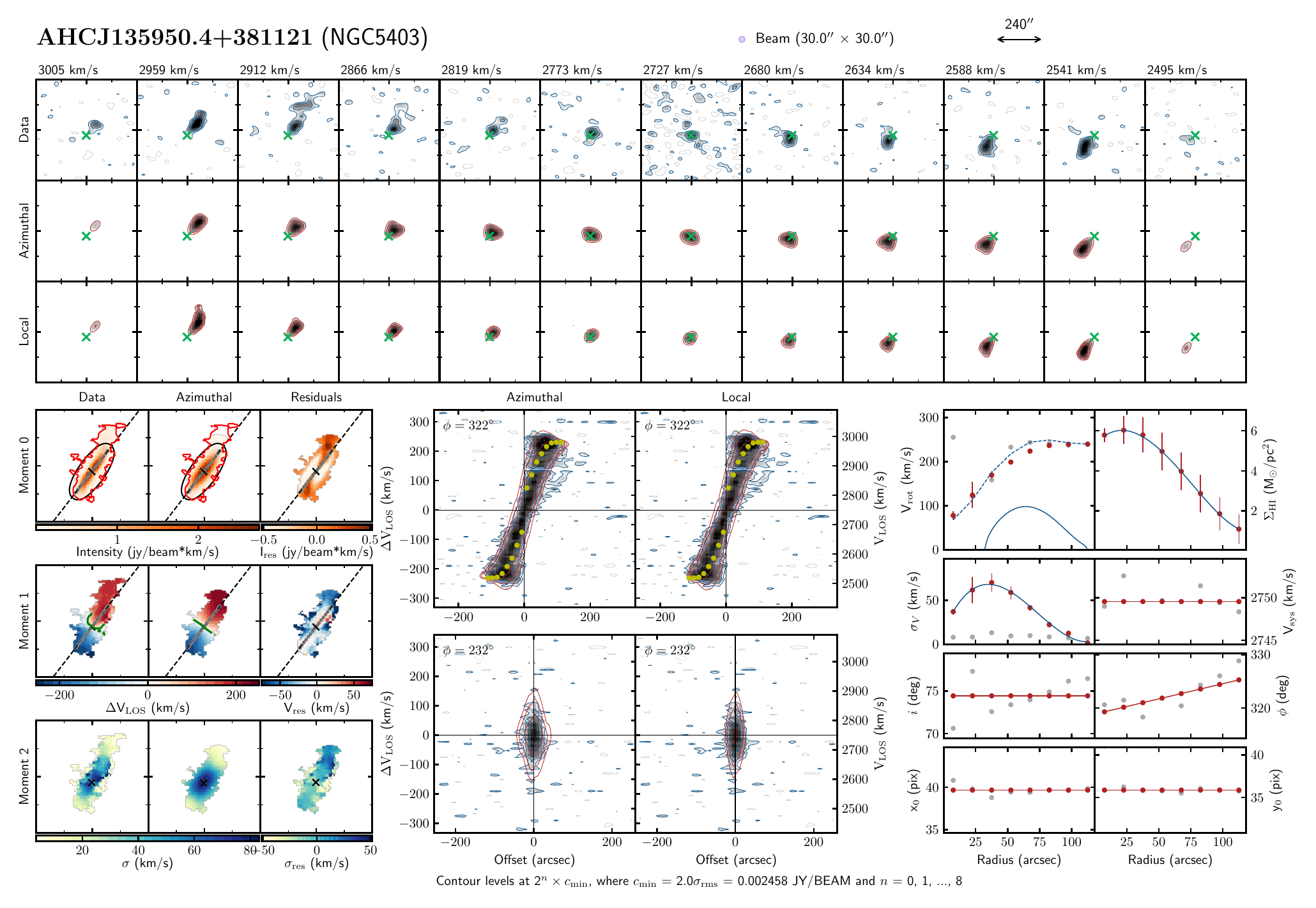}
    \caption{\bbarolo\ output for an $A = 3$ kinematic model. Panels are as in Fig.~\ref{fig_a1_plot1}. The galaxy should have been modeled with the edge-on pipeline, as can be seen from the channel maps of the azimuthally normalized model and the unusually high velocity dispersion in the center.}
    \label{fig_a3_plot2}
\end{figure*}

On the other hand, the $A = 3$ models shown in Figs.~\ref{fig_a3_plot1} and \ref{fig_a3_plot2} are not acceptable as they are. For the first one of these (AHCJ012311.4+333150, NGC\,496) \bbarolo\ incorrectly moved the kinematic center away from the optical center, which leads to a wrong determination of the rotation curve shape. Similarly, the second $A = 3$ model shown in Fig.~\ref{fig_a3_plot2} (AHCJ135950.4+381121, NGC\,5403) is not acceptable because the standard pipeline was used despite the galaxy being edge-on. This happened because the galaxy's inclination is underestimated in the PGC catalog (${\sim}\SI{77}{\deg}$), likely due to a strong bulge, which is visible in our DESI Legacy images. The typical signs of this error are the butterfly-like patterns that are visible in the channel maps of the azimuthally normalized model, but not in those of the data, and the unusually large central velocity dispersion ($>\SI{50}{\kilo\meter\per\second}$).

\begin{figure*}[p]
    \centering
    \includegraphics[width = \textheight - 40pt, angle = 90]{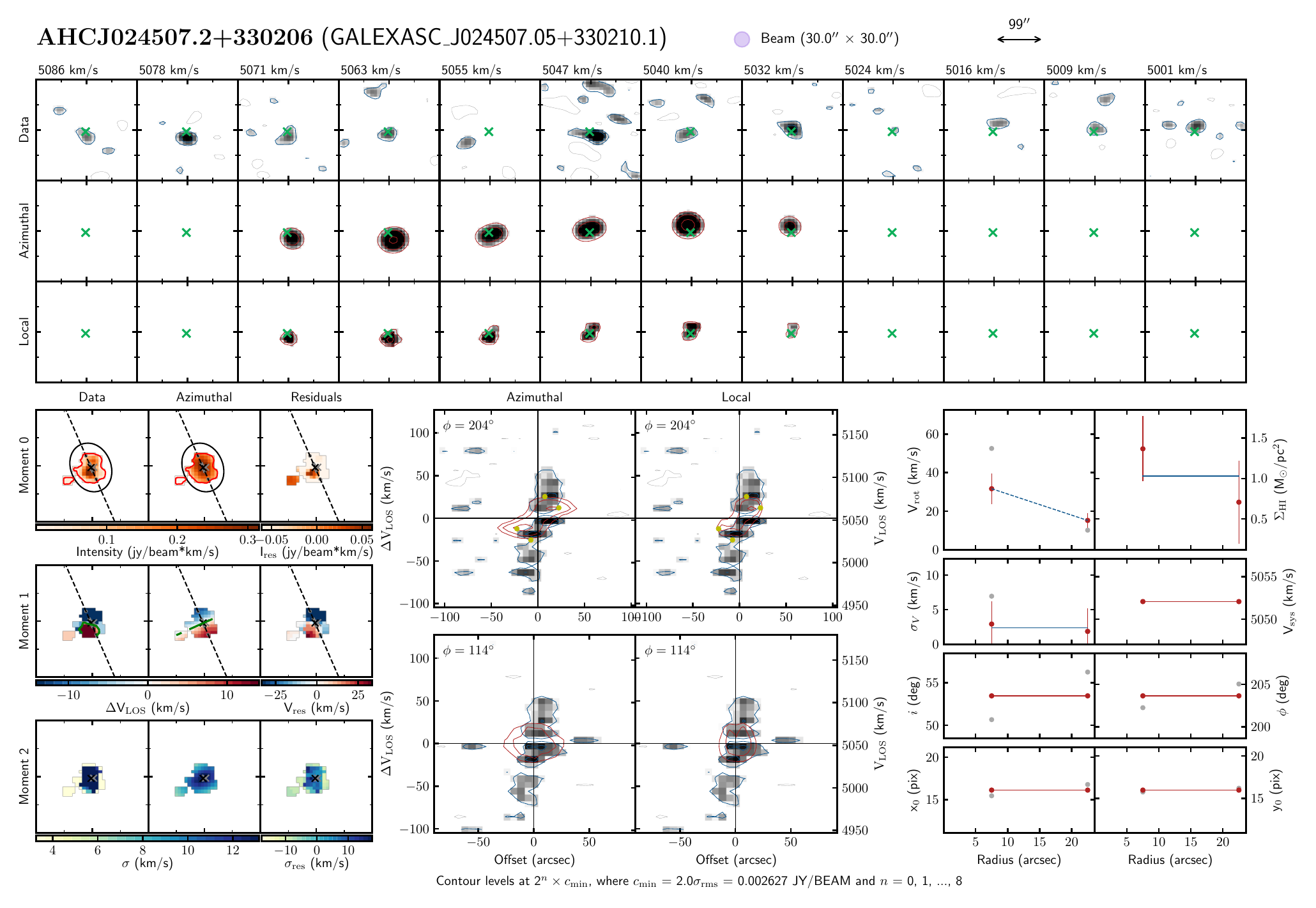}
    \caption{\bbarolo\ output for an $A = 4$ observation. Panels are as in Fig.~\ref{fig_a1_plot1}. The galaxy is rejected for kinematic modeling because it is resolved with less than two beams along the major axis.}
    \label{fig_a4_plot}
\end{figure*}

Finally, Fig.~\ref{fig_a4_plot} shows the overview plots for a galaxy where the data does not allow for the construction of a reliable model. In this case, the galaxy (AHCJ024507.2+330206) is resolved with less than two beams along the major axis.

\end{appendix}

\end{document}